\documentclass[epj]{svjour}
\usepackage{amssymb}
\usepackage{enumitem}
\usepackage{amsmath}
\usepackage{hyperref}
\usepackage{graphicx}
\begin{document}
\title{Modelling LARES temperature distribution and thermal drag}
\author{Phuc H. Nguyen \inst{1} \thanks{\emph{Email:} phn229@physics.utexas.edu} \and Richard Matzner\inst{1}
\thanks{\emph{Email:} matzner2@physics.utexas.edu}%
}                     
\institute{Department of Physics and Texas Cosmology Center, The University of Texas, Austin, TX 78712, USA
}
\date{Received: date / Revised version: date}
%
\abstract{
 The LARES satellite, a laser-ranged space experiment to contribute to geophysics observation,  and to measure the general relativistic Lense-Thirring effect, has been observed to undergo an anomalous along-track orbital acceleration of -0.4 pm/s$^2$ (pm := picometer). This thermal ``drag'' is not surprising; along track thermal drag has previously been observed with the related LAGEOS satellites (-3.4 pm/s$^2$). It is hypothesized that the thermal drag is principally due to anisotropic thermal radiation from the satellite's exterior.\\
We report the results of numerical computations of the along-track orbital decay of the LARES satellite during the first 126 days after launch. The results depend to a significant degree on the visual and IR absorbance $\alpha$ and emissivity $\epsilon$ of the fused silica Cube Corner Reflectors. We present results for two values of $\alpha_{IR}$ = $\epsilon_{IR}$: 0.82, a standard number for ``clean'' fused silica; and 0.60, a possible value for silica with slight surface contamination subjected to the space environment.\\
The heating and the resultant along-track acceleration depend on the plane of the orbit, the sun position, and in particular on the occurrence of eclipses, all of which are functions of time. Thus we compute the thermal drag for specific days. We compare our model to observational data, available for a 120-day period starting with the 7th day after launch, which shows the average acceleration of -0.4 pm/s$^2$. With our \textit{model} the average along-track thermal drag over this 120-day period for CCR $\alpha_{IR}$ = $\epsilon_{IR}$ = 0.82 was computed to be -0.59 pm/s$^2$. For CCR $\alpha_{IR}$ = $\epsilon_{IR}$ = 0.60 we compute -0.36 pm/s$^2$.\\
LARES consists of a solid spherical tungsten sphere, into which the CCRs are set in colatitude circles.  Our calculation models the satellite as 93 isothermal elements: the tungsten part, and each of the 92 Cube Corner Reflectors. The satellite is heated from two sources: sunlight and Earth's infrared (IR) radiation. We work in the fast-spin regime, where CCRs with the same colatitude can be taken to have the same temperature. Since all temperature variations (temporal or spatial) are expected to be small, we linearize the Stefan-Boltzmann law and, taking advantage of the linearity, we perform a Fourier series analysis. The variations are indeed small, validating our Fourier analysis.
} 
\maketitle
\tableofcontents
\section{Introduction}
LARES is a space mission sponsored by the Italian and European space agencies, which consists of a passive satellite tracked by the global network of laser ranging stations of the International Laser Ranging Service. The satellite is a ball of tungsten 18.2 cm in radius and weighing about 400 kg, with 92 fused silica ({\it Suprasil}) Cube Corner Reflectors (CCRs) covering the surface. The orbit is nearly circular, with a perigee of 1450 km and an orbital inclination of 69.5 degree. It is similar in structure (but smaller and denser) to the related satellite LAGEOS. LARES consists of a solid tungsten core, whereas the LAGEOS satellites have a cylindrical brass core surrounded by an aluminum shell consisting of two joined hemispheres, into which the CCRs are set.\\
One of LARES's scientific targets is to verify the Lense-Thirring effect \cite{Thirring(1918),Thirring(1921),LenseThirring(1918)} (also known as frame-dragging), a prediction of the theory of General Relativity, using the technique of laser ranging with an accuracy of about 1 \% \cite{Ciufolini(1986),Ciufolini(1989),Ciufolini(1993),Ciufolini(1994),CiufoliniParis(2012)}. Frame-dragging is the phenomenon by which a spinning mass drags the local inertial frame in the direction of the rotation. This induces a precession of the plane of an orbiting test body \cite{CiufoliniSindoni(2012)}. The {\it ascending node} of an orbit around the Earth drifts eastward. In the weak field slow rotation limit appropriate to orbits around the Earth, this precession is given by given by:
\begin{equation}
\dot{\vec{\Omega}}_{Lense-Thirring} = \frac{2\vec{J}}{a^{3}(1-e^{2})^{3/2}}
\end{equation}
For an object in low orbit around Earth, in particular, the formula above yields a precession of the order of a tens or hundreds of milliarcsecond per year, corresponding to meters per year motion of the node.  Studies using Earth gravity field from the GRACE mission together with the laser ranged satellites LAGEOS and LAGEOS 2  have observed the node dragging to approximately 10 \% \cite{CiufoliniPavlis(2004)}. Due to the smallness of the Lense-Thirring effect, experimental detection is a real challenge and it is crucial to be able to model all the experimental sources of errors as accurately as possible \footnote{The frame dragging leads also to a related effect, precession of the direction of a gyroscope: the direction of the gyroscope which initially points to a specific celestial coordinate (eg toward a particular QSO), deviates from this direction at the rate 
\begin{equation}\label{GPBprecession}
\vec{\omega}_{gyro} = \frac{GI}{c^{2}r^{3}}\left[\frac{3\vec{r}}{r^{2}}\left(\vec{S}\cdot\vec{r}\right)-\vec{S}\right]
\end{equation}
where $\vec{\omega}_{gyro}$ is the gyroscope precession, $I$ is the moment of inertia of Earth, $\vec{r}$ is the position of the gyroscope, and $\vec{S}$ is Earth's angular velocity. At the 642 km altitude of the GP-B experiment \cite{Everitt(2011)}, equation (\ref{GPBprecession}) gives a precession rate of $-39.2$ milliarc-second per year, where the minus sign indicates that the precession is Eastward (\cite{Everitt(2011)} Fig. 1). GPB observed a frame-dragging of $-37.2 \pm 7.2$ milliarc second per year, about a 19\% quoted error.}.\\
There are many contributions to the nodal precession rate, both gravitational and non-gravitational. Many of these perturbations are orders of magnitudes larger than the Lense-Thirring effect and not all of them are well-modelled. Precession due to Newtonian effects arises from the fact that Earth is not a perfect sphere. Non-gravitational effects include thermal and residual-gas effects.\\
In the rest of this paper, we will focus on the measure of one of the most important non-gravitational sources of error: thermal drag. This is the net deceleration that the satellite undergoes due to a nonuniform temperature distribution (and therefore a nonuniform thermal radiation at the surface). In the case of LAGEOS, this net force explains that satellite's anomalous along-track acceleration of about -3.4 cm/s$^{2}$. For LARES's first 126 days in orbit, its average drag over the last 120-day period was measured as -0.4 pm/s$^2$ \cite{CiufoliniMaztner(2012)}. We will model the temperature of the LARES satellite with the following two assumptions (justified below): (1) the temperature variation inside the metal part of the satellite or within a single CCR is ignorable compared to the temperature variation between the metal and any CCR or between different CCRs, and (2) the satellite spin is sufficiently fast that CCRs of the same colatitude (with respect to the spin axis) have the same temperature.\\
The rest of the paper is organized as follows. In Section \ref{Sec:Structure}, we describe the structure of the satellite in detail. In Section \ref{Sec:Biot}, we estimate in a simple way the temperature variation within the tungsten, and within each CCR, thus justifying the assumption that temperature variations inside each CCR and the metal are ignorable. In Section \ref{Sec:ViewFactors}, we compute the so-called view factors of the CCR cavity, which are necessary ingredients to handle the radiative heat transfer inside the cavity. In Section \ref{Sec:Geometry}, we describe the geometry of the orbital plane, the spin orientation, and the solar eclipse. In Section \ref{Sec:Yarkovsky}, we give an overview of the satellite's thermal response and thermal drag. In Section \ref{Sec:Analysis}, we present the results of our calculation.  In Section \ref{Sec:Conclusion}, we summarize the main findings of our study. In Appendix \ref{App:ViewFactors}, we review the main elements needed to handle radiative heat exchange problems, in particular the formalism of view factors and the radiosity method. In Appendix \ref{App:Constants}, we list the numerical values of all the constants used throughout the paper. Finally, in the two last appendices (Appendix \ref{App:SpinningSphere} and \ref{App:ClebschGordan}), we present two analytical computations to find the temperature inside a metal sphere bathed in sunlight. Appendix \ref{App:SpinningSphere} focuses on the case of a spinning sphere without orbital motion. While these computations are not directly related to the main points of the paper, they employ several of the tools also used in the paper (especially the Fourier series technique), and can be of interest to future researchers in the field of radiative heat exchange. Finally, Appendix \ref{App:ClebschGordan} considers a stationary, nonspinning sphere, and discusses a possible treatment to go beyond the linearized boundary conditions.

\section{Satellite structure}\label{Sec:Structure}
\begin{figure}
$$
\begin{array}{cc}
  \includegraphics[width=8cm]{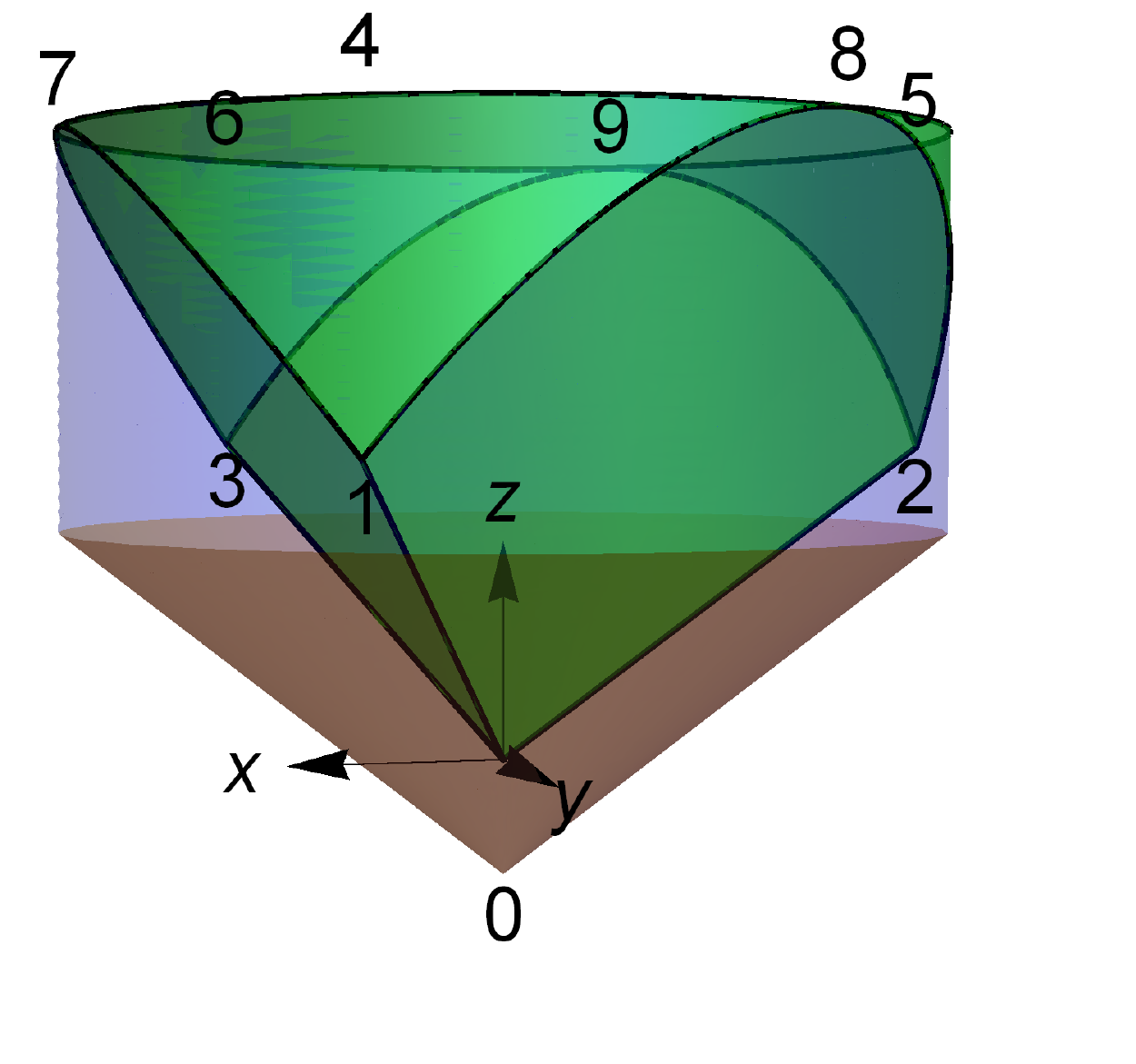}
\end{array}
$$
\caption{The shape of the CCR.}
\label{CCRshape}
\end{figure}

In this section, we describe the arrangement of the CCRs on the satellite, and the shape of each CCRs \cite{Paolozzi(2009),Paolozzi(2011)}. There are 92 CCRs on LARES, and they are arranged into 10 rows. We give below the number of CCRs per row ($N_{I}$ where $I$ labels the row) and the colatitude $\theta_{I}$ of the row. The CCRs belonging to a given row are equally spaced.
\begin{center}
\begin{tabular}{|l|l|l|l|l|l|}
  \hline
  Row & $n_{I}$ & $\theta_{I}$ & Row & $n_{I}$ & $\theta_{I}$ \\
  \hline
  I & 1 & 0 deg & -V & 16 & 100 deg \\
  II & 5 & 20 deg & -IV & 14 & 120 deg \\
  III & 10 & 40 deg & -III & 10 & 140 deg \\
  IV & 14 & 60 deg & -II & 5 & 160 deg \\
  V & 16 & 80 deg & -I & 1 & 180 deg \\
  \hline
\end{tabular}
\end{center}
The shape of each CCR is depicted in fig. \ref{CCRshape}. The CCR is a right triangular pyramid intersected by a cylinder of radius $R$. The CCR fits inside a cylindrical tungsten cavity with a conical bottom. We will denote by $d$ ($\approx$ 5 mm) the distance between the tip of the CCR and the bottom of the conical floor of the cavity.
For the sake of visualization, we have labelled a few points on fig. \ref{CCRshape} and given the coordinates of those points in the table below. The origin of the coordinate system was chosen to be at the tip of the CCR.
\begin{center}
\begin{tabular}{|l|l|l|l|}
  \hline
  Point & $(\frac{x}{R},\frac{y}{R},\frac{z}{R})$ & Point & $(\frac{x}{R},\frac{y}{R},\frac{z}{R})$\\
  \hline
  $0$ & $(0,0,-\frac{d}{R})$ & $5$ & $(-1,0,\sqrt{2})$ \\
  $1$ & $(\frac{1}{2},\frac{\sqrt{3}}{2},\frac{1}{\sqrt{2}})$ & $6$ & $(\frac{1}{2},-\frac{\sqrt{3}}{2},\sqrt{2})$ \\
  $2$ & $(-1,0,\frac{1}{\sqrt{2}})$ & $7$ & $(1,0,\sqrt{2})$ \\
  $3$ & $(\frac{1}{2},-\frac{\sqrt{3}}{2},\frac{1}{\sqrt{2}})$ & $8$ & $(-\frac{1}{2},\frac{\sqrt{3}}{2},\sqrt{2})$ \\
  $4$ & $(\frac{1}{2},\frac{\sqrt{3}}{2},\sqrt{2})$ &   $9$ & $(-\frac{1}{2},-\frac{\sqrt{3}}{2},\sqrt{2})$ \\
  \hline
\end{tabular}
\end{center}

\section{A few simple estimations}\label{Sec:Biot}
In this section, we justify the assumption of uniform temperature inside the metal and each CCR by estimating the temperature variation within the metal and within a CCR. First suppose the satellite to be a simple tungsten ball in radiative exchange with empty space, and heated by sunlight (in this section we will ignore Earth IR radiation). To compute the power absorbed by the ball, notice that the heat absorbed by each area element $dA$ on the sphere is proportional to the cosine of the angle between the normal to $dA$, i.e. the cosine of the colatitude of $dA$. Integrating over the illuminated hemisphere then yields:
\begin{equation}\label{Q}
Q = \alpha_{W,vis} \Phi R_{sat}^{2} \int_{0}^{2\pi} \int_{0}^{\pi/2} \cos{\theta} \sin{\theta}d\theta d\phi
\end{equation}
where $R_{sat}$ is the radius of the LARES satellite, $\alpha_{W,vis}$ is the absorptivity of tungsten in the visible and $\Phi$ is the solar constant. The values of all constants appearing in this paper are tabulated in Appendix \ref{App:Constants}. Evaluating the integral above, we find:
\begin{equation}
Q = \alpha_{W,vis}\Phi\pi R_{sat}^{2}
\end{equation}
i.e. the heat absorbed is proportional to the \textit{cross-section} $\pi R_{sat}^{2}$.\\
By Fourier's law of heat conduction, $Q$ is roughly the product of the temperature gradient, the cross-section area through which heat is conducted, and the thermal conductivity $\kappa_{W}$ of tungsten:
\begin{equation}
Q \approx \kappa_{W}\frac{\Delta T}{R_{sat}}\pi R_{sat}^{2}
\end{equation}
The two equations above yield
\begin{equation}
\Delta T \approx \frac{\alpha_{W,vis}\Phi R_{sat}}{\kappa_{W}} \approx \mathrm{1 K}
\end{equation}
On the other hand, conservation of energy reads:
\begin{equation}
\epsilon_{W,IR} \sigma T_{W}^{4} 4 \pi R_{sat}^{2} = \alpha_{W,vis} \Phi \pi R_{sat}^{2},
\end{equation}
where $\epsilon_{W,IR}$ is the emissivity of tungsten in the IR, $\sigma$ is the Stefan-Boltzmann constant, and $T_{W}$ is the metal's temperature. From this equation, we find:
\begin{equation}
T_{W} = \left(\frac{\alpha_{W,vis}\Phi}{4\epsilon_{W,IR}\sigma}\right)^{1/4} \approx \mathrm{443.6 K}
\end{equation}
Thus, the temperature variation $\Delta T$ is only about $0.2$ percent of the average temperature. We can therefore take the metal to be at a uniform temperature to an excellent degree of accuracy.\\
Next, we turn to the estimation of the temperature variation inside a CCR. For simplicity, we can model the CCR as a cone with half-angle $\frac{\pi}{4}$ inside a cylindrical cavity of radius and height $R$, and such that the base of the cone is flush with the satellite's surface. The CCR is in radiative heat exchange with the metal wall of the cavity, which is assumed to be at constant temperature $T_{W}$ as previously found. We will assume no conduction takes place from the metal to the CCRs \footnote{In reality, the CCRs are not in contact with the metal, but are held in place by Teflon spacers, which are covered at the surface by a tungsten retainer ring. Teflon is a very good thermal insulator, so we assume the CCRs have only radiative contact with the metal. Moreover, we will ignore the effect of the retainer rings; the temperature of a copy of LAGEOS 2's retainer rings was measured in a 2006 laboratory test. The retainer rings were found to have closely the same temperature as the body of the satellite \cite{Paolozzi(2015)}. Hence the retainer rings should not have any noticeable effect on the thermal drag.}.\\
An exercise in radiative heat exchange (see Appendix \ref{App:ViewFactors}) shows that the net power transferred from the metal to the CCR is given by:
\begin{equation}
P_{net} = \epsilon_{eff}A_{gl}\sigma(T_{W}^{4}-T_{gl}^{4})
\end{equation}
where $A_{gl}=\sqrt{2}(\pi R^{2})$ is the surface area of the glass wall, $T_{gl}$ is the average CCR temperature, and $\epsilon_{eff}$ is an effective emissivity which is solely determined by $\epsilon_{gl,IR}$, $\epsilon_{W,IR}$ as well as the geometry of the cavity. To compute $\epsilon_{eff}$, we need the so-called view factors between isothermal elements lining the cavity $F_{i \rightarrow j}$, which represent the fraction of radiation leaving isothermal element $i$ which strikes element $j$. In our case, we only have 2 isothermal elements: the glass and the metal. Moreover, since the CCR is a convex surface, two of the view factors are trivial:
\begin{equation}
F_{gl,gl} = 0
\end{equation}
and
\begin{equation}
F_{gl,W} = 1
\end{equation}
The other two view factors follow from the surface area of the glass ($A_{gl}$) and the metal ($A_{W}$) lining the cavity:
\begin{equation}
A_{gl} = \sqrt{2}\pi R^{2}
\end{equation}
\begin{equation}
A_{W} = 3\pi R^{2}
\end{equation}
Using the identities (\ref{reciprocity}) and (\ref{sumrule}) for view factors, we find:
\begin{equation}
F_{W,gl} = \frac{\sqrt{2}}{3}
\end{equation}
and
\begin{equation}
F_{W,W} = 1 - \frac{\sqrt{2}}{3}
\end{equation}
Finally, the effective emissivity is given in terms of the view factors by:
\begin{equation}
\epsilon_{eff} = \bigg( \frac{1}{\epsilon_{gl,IR}} + \frac{1-\epsilon_{W,IR}}{\epsilon_{W,IR}} F_{W,gl} \bigg)^{-1}
\end{equation}
Specifically, in our case:
\begin{equation}
\epsilon_{eff} = \bigg( \frac{1}{\epsilon_{gl,IR}} + \frac{1-\epsilon_{W,IR}}{\epsilon_{W,IR}}\frac{\sqrt{2}}{3} \bigg)^{-1}
\end{equation}
with $\epsilon_{gl,IR}$ the emissivity of glass in the IR. This net heat is then radiated into space by the base of the cone, which faces into space:
\begin{equation}
P_{net} = \epsilon_{gl,IR} \sigma T_{gl}^{4} \pi R^{2}
\end{equation}
We can now solve for the average glass temperature \footnote{We are assuming that the CCR is not illuminated by the Sun in this simple example.}:
\begin{equation}
T_{gl} = \bigg(1+\frac{\epsilon_{gl,IR}}{\sqrt{2}\epsilon_{eff}}\bigg)^{-1/4} T_{W} \approx 291.8 \ \mathrm{K}
\end{equation}
Using Fourier's Law, we can approximate the heat flow in the glass just below the surface to:
\begin{equation}
P_{net} = \kappa_{gl}\frac{\Delta T}{R}\pi R^{2}
\end{equation}
where $\kappa_{gl}$ is the thermal conductivity of glass. The temperature variation inside the CCR is then:
\begin{equation}
\Delta T = \frac{\epsilon_{eff}A_{gl}\sigma(T_{W}^{4}-T_{gl}^{4})}{\kappa_{gl}\pi R} \approx \mathrm{3.84 K}
\end{equation}
This represents 1.3 percent of the average CCR temperature. This result is consistent with experimental (ground based) measurements of the CCR temperature distribution \cite{Paolozzi(2015)}.
\section{CCR view factors}\label{Sec:ViewFactors}
We will assume that only radiative heat transfer takes place between the CCR and the tungsten, and no heat is conducted. In order to use the radiosity method, we will need to write down the view factors and the effective emissivity for the geometry depicted in fig. \ref{CCRshape}. First, the total metal surface area lining the cavity is:
\begin{equation}
A_{W} = 2\pi R(\sqrt{2}R-2d) + \pi R \sqrt{R^{2}+9d^{2}}
\end{equation}
The total glass surface area lining the cavity is:
\begin{equation}
A_{gl} = (\sqrt{3}\pi + 2\sqrt{2}\pi -3\sqrt{6})R^{2}
\end{equation}
Notice that, like in Section \ref{Sec:Biot}, the CCR is still a convex surface, implying that two of the four view factors still vanish:
\begin{equation}
F_{gl,W} = 1
\end{equation}
\begin{equation}
F_{gl,gl} = 0
\end{equation}
The other two view factors can be found using the areas $A_{W}$ and $A_{gl}$, and the relations (\ref{reciprocity}) and (\ref{sumrule}):
\begin{equation}
F_{W,W} = \frac{(3\sqrt{6}-\sqrt{3}\pi)R^{2}-4\pi Rd + \pi R\sqrt{R^{2}+9d^{2}}}{2\pi R(\sqrt{2}R-2d) + \pi R \sqrt{R^{2}+9d^{2}}}
\end{equation}
\begin{equation}
F_{W,gl} = \frac{(\sqrt{3}\pi+2\sqrt{2}\pi-3\sqrt{6})R^{2}}{2\pi R(\sqrt{2}R-2d) + \pi R \sqrt{R^{2}+9d^{2}}}
\end{equation}
\section{Orbital geometry}\label{Sec:Geometry}
\begin{figure}
$$
\begin{array}{cc}
  \includegraphics[width=10cm]{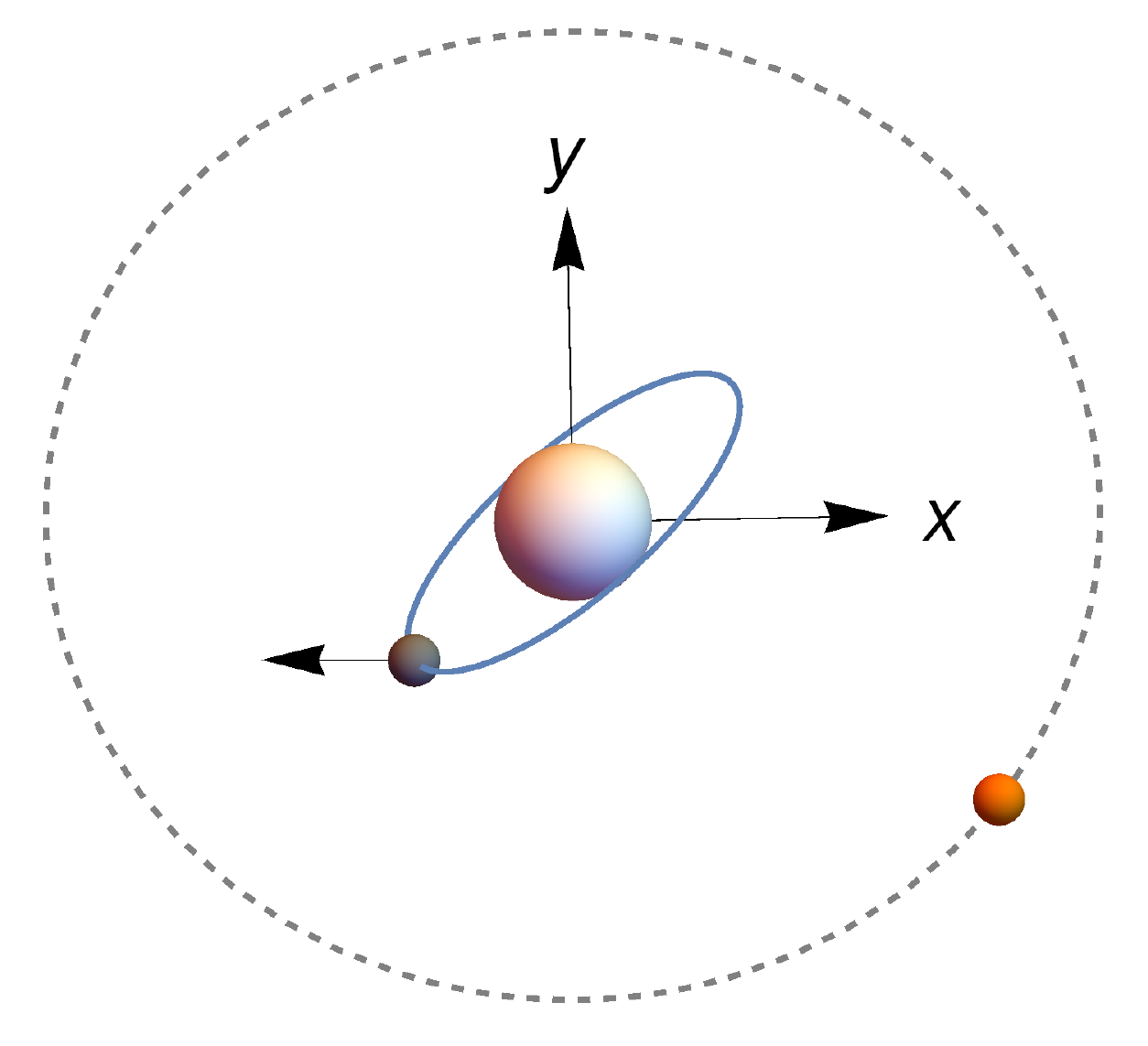}
\end{array}
$$
\caption{The situation at $t = 0$ and $k = 0$, as viewed directly above the North Pole of Earth. The sphere centered at the origin represents the Earth. The coordinate axes form the celestical coordinate system. The orange sphere is the Sun. The dashed gray circle is the Sun orbit through 1 year. The gray sphere is LARES, with the vector representing the spin orientation. The blue ellipse is the projection of LARES’s trajectory for k = 0 (i.e. day zero, launch day) on the x-y plane. The radial distances from the Earth's center are not drawn to scale. However all angular information in the sky is drawn accurately.}
\label{OrbitalGeometry}
\end{figure}

In this section, we describe the geometry of the orbital motion and the spin axis (see fig. \ref{OrbitalGeometry} for an illustration of the orbital geometry on the day of launch). The orbit is nearly polar, highly circular with inclination $i = 69.5$ degrees, which we take as $70$ degrees in our model. Due to the Earth's quadrupole moment, the orbital plane precesses westward at a rate of $1.7$ degree/day. This value, and all values for the orbital parameters for the first 126 days are taken from \cite{CiufoliniNeumayer(2012)}. The satellite was successfully launched in orbit on February 13, 2012 at 10:00 UTC from the European Space Agency's spacesport in Kourou, French Guyana \cite{PaolozziCiufolini(2013)}. To describe the orbit, we will work in celestial coordinate system \footnote{i.e. Cartesian coordinate system, with the Earth's equatorial plane as the x-y plane, and the x-axis pointing toward the vernal equinox.}. The unit vector $\hat{r}_{Sun}$ pointing from Earth to Sun $k$ days after launch is:
\begin{equation}
\hat{r}_{Sun}{(k)} = \begin{pmatrix}
\cos{(2\pi(k-37)/365)}\\
\cos{\beta}\sin{(2\pi(k-37)/365)}\\
\sin{\beta}\sin{(2\pi(k-37)/365)}
\end{pmatrix}
\end{equation}
where $\beta$ = 23.2 degree is the obliquity of the Sun's orbital plane (i.e. the angle this plane makes with the Earth equatorial plane), and 37 is the number of days from the date of launch of LARES to the vernal equinox. We will assume that for any given day, the position of the Sun in the sky and $\Omega$ stay fixed (i.e. they increment discretely from one day to the next). Also, the unit vector $\hat{r}_{sat}$ from the center of Earth to the center of the satellite is:
\begin{equation}
\hat{r}_{sat}{(t,k)} = \begin{pmatrix}
-\sin{\Omega}\cos{i}\sin{\omega_{o} t}+\cos{\Omega}\cos{\omega_{o} t}\\
\cos{\Omega}\cos{i}\sin{\omega_{o} t}+\sin{\Omega}\cos{\omega_{o} t}\\
\sin{i}\sin{\omega_{o} t}
\end{pmatrix}
\end{equation}
where $\omega_{0}$ is the satellite's orbital frequency, and
\begin{equation}
\Omega{(k)} = (220-1.7k)\frac{\pi}{180}
\end{equation}
is the longitude of the ascending node on day $k$ \footnote{For a geocentric orbit such as LARES, the longitude of the ascending node (also known as the right ascension of the ascending node or RAAN) is the angle, measured eastward, from the vernal equinox to the ascending node. The ascending node is defined to be the intersection of LARES's orbit with the Earth's equatorial plane, where LARES is heading north.}. The time $t$ will be taken to range in $t \in [0,T]$, i.e. one orbital period, with the satellite at the longitude of ascending node at $t=0$.\\
The spin orientation and spin frequency have been measured and the results are reported in \cite{Kucharski}. The spin orientation is at $RA = 12^{h}22^{m}48^{s}$ ($RMS=49^{m}$) and $Dec=-70.4$ degrees ($RMS=5.2$ degrees). For simplicity, we will take the spin direction to be (in celestial coordinates):
\begin{equation}
\hat{S} = \begin{pmatrix}
-\cos{i}\\
0\\
-\sin{i}
\end{pmatrix}
\end{equation}
where $i = 70$ degrees is the orbital inclination of LARES. The above choice for $\hat{S}$ is within the margin of error for the RA and Dec. Next, the unit vector $\hat{r}_{CCR}$ pointing from LARES's center to a CCR at colatitude $\theta_{I}$  is:
\begin{equation}
\hat{r}_{CCR}{(t,\theta_{I})} = \begin{pmatrix}
-\sin{i}\sin{\theta_{I}}\cos{\omega t}-\cos{i}\cos{\theta_{I}}\\
\sin{\theta_{I}}\sin{\omega t}\\
\cos{i}\sin{\theta_{I}}\cos{\omega t}-\sin{i}\cos{\theta_{I}}
\end{pmatrix}
\end{equation}
where $\omega$ is the spin frequency. Notice that the time dependence in the vector above is entirely due to the spinning of the satellite. The spin frequency actually decreases over time due to interaction with the Earth's magnetic field, as (\cite{Kucharski}):
\begin{equation}
\omega{(k)} = (\mathrm{0.546 rad/s})\exp{-0.00322509k}
\end{equation}
For $k=0$, the spin angular frequency is about 600 times the orbital frequency. For $k \approx 100$, the spin frequency has decreased to about 400 times the orbital frequency. Thus, we can see that during the first 126 days, we remain safely inside the fast-spin regime.

\section{CCR heating and thermal drag}\label{Sec:Yarkovsky}
The physical concepts that lead to along-track thermal drag for LAGEOS and LARES style satellite (passive metal satellites with CCR retroreflectors) were developed over a number of years by \cite{RubincamWeiss(1985),Rubincam(1987),Rubincametal(1987),Rubincam(1988),Afonsoetal(1989),Farinellaetal(1990),Rubincam(1990),Rubincametal(1997)} and incorporated into Slabinski's definitive study of the along track thermal drag on LAGEOS \cite{Slabinski(1996)}. Our analysis is based on these references, though with technical differences; we do Fourier analyses while Slabinski did direct time integration, for instance. In this Section we discuss the two concepts most crucial to understand the LARES along-track thermal drag. Our approach will be to linearize the thermal equations and develop our study in terms of Fourier analysis. This means that the heating due to Earth IR and that due to solar heating can be treated, at least conceptually, separately. In our actual computations we sum all contributions.\\
As we have noted, the tungsten core of LARES has very high conductivity, and small temperature differences (which we ignore) across its surface, regardless of the heating source. However the CCRs are {\it glass} which is a good heat insulator, has a substantial heat capacity, and is an efficient infrared radiator. These properties mean that the CCRs have substantial thermal inertia. Thus the CCRs heat gradually when exposed to a heating source, and cool gradually when the source is removed.\\
\subsection{Earth Infrared CCR heating and thermal drag}\label{Sec:IRYarkovsky}
\begin{figure}
$$
\begin{array}{cc}
  \includegraphics[width=10cm]{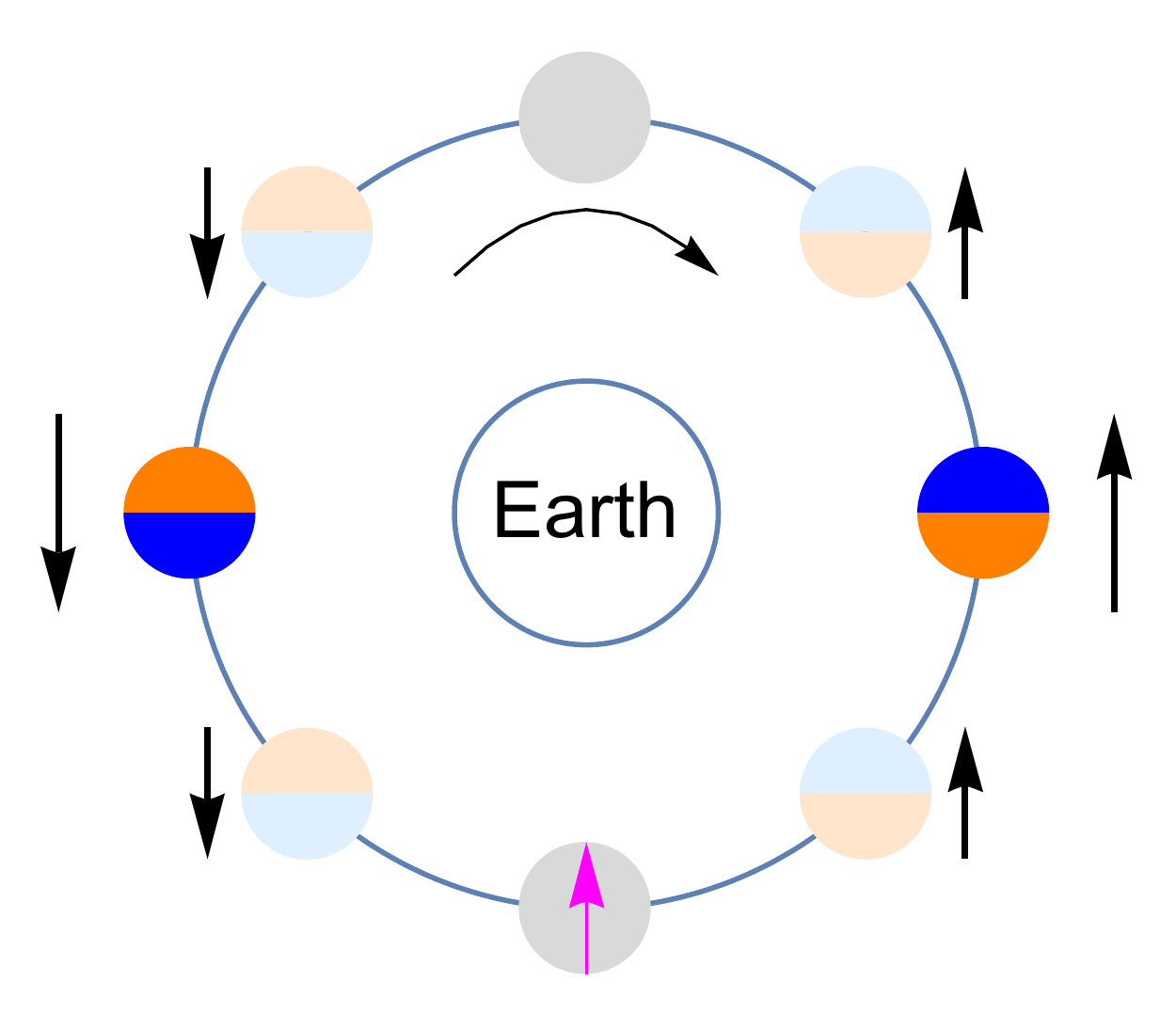}
\end{array}
$$
\caption{Illustration of the Yarkovsky effect due to Earth IR radiation. The satellite is depicted at 8 different positions along its orbit, and its spin axis is the arrow in magenta at the bottom. The direction of motion of the satellite is clockwise. The hotter hemisphere is depicted in orange, and the cooler hemisphere is depicted in blue. A sharper color contrast between the two hemisphere indicates a greater thermal anisotropy. When there is no anisotropy, the satellite is depicted in gray. The length of the black arrow indicates the magnitude of the axial force. For simplicity, we depict the situation where the thermal lag {\it phase angle} is $\pi/4$.}
\label{Yarkovsky}
\end{figure}

The thermal drag from Earth-IR heating is due to the Earth Yarkovski effect \cite{Rubincam(1988)} (see fig. \ref{Yarkovsky} for an illustration of this effect). The simplest example has the spin axis in plane of orbit. Maximum hemispheric IR absorption occurs when one ”pole”  points toward Earth IR source. Because the CCR -glass- is opaque to IR, with low thermal conductivity, CCRs near the heated pole warm slowly, so the along axis thrust builds up after the satellite has moved from the the position of maximum hemispheric IR absorption. (In the fast spin case we consider, the temperature distribution is axisymmetric around the spin axis.) The along-track thermal drag (it is always a drag) maximizes after some {\it phase angle} in the orbit. The maximum effect occurs twice per orbit.\\
If the spin axis is out of orbital plane there is less differential heating, and the need to project resulting (smaller) thrust along orbit reduces the along-orbit thermal drag further. But it is always a drag.

\subsection{The eclipse}
For certain configurations of orbital plane and sun position, the satellite will undergo (solar) eclipses as it passes behind the Earth, i.e. into the shadow of the Earth, and is temporarily shielded from direct solar radiation.\\
When there are {\it no} eclipses, in the fast spin regime solar radiation has no effect on the thermal drag averaged over an orbit. This is because the  spin is parallel propagated, so always points to the same celestial coordinates, and for one orbit, the solar celestial position is very closely constant also. Thus the same hemisphere of the satellite is heated throughout the entire orbit. The consequent spin-axis thrust is constant in size, and constant in direction (along the fixed spin axis) throughout the orbit. However, when dotted into the orbital velocity (to obtain the along-track acceleration), it is alternately a positive acceleration, and a drag on the opposite side of the orbit, so that the along track thermal drag arising from this source averages out over one orbit.\\
When there is an eclipse, however, the heating is removed for part of the orbit. If we incorrectly assume that the thermal thrust immediately disappears during the eclipse, then the along track acceleration during the eclipse is no longer available to cancel the contribution from the other side of the orbit. This leads to a nonzero net along-track acceleration. This net acceleration can be either positive, or negative. In his LAGEOS studies, Slabinski \cite{Slabinski(1996)} found that the size of the effect could be as large as the Earth-IR induced thermal drag, and because of the possibility of either sign depending on the spin/orbital plane/sun geometry, could swing the thermal drag to twice the noneclipse thermal drag, or  zero, consistent with LAGEOS observations.\\
In both LARES and LAGEOS, the thermal inertia of the system means that the thermal thrust does not disappear immediately upon entering the eclipse, but decays slowly, and recovers gradually when the satellite emerges from the eclipse. But the principle is as explained in the previous paragraph. Also note that because of the lower orbit of LARES, it spends a much larger fraction of its time in eclipse than does LAGEOS.\\
We will need to compute the time of eclipse entrance $t_{entrance}$ and eclipse exit $t_{exit}$. To do this, we compute the perpendicular distance $d{(t,k)}$ from the satellite to the Earth-Sun axis:
\begin{equation}
d{(t,k)} = a |\hat{r}_{sat}{(t,k)} \times \hat{r}_{Sun}{(k)}|
\end{equation}
Then $t_{entrance}$ and $t_{exit}$ are solutions to the equation (see footnote \ref{REfootnote} below):
\begin{equation}
d{(t,k)} = R_{E}
\end{equation}
However, in general there will be 4 solutions. To see this, one can visualize a cylinder with the Earth-Sun axis for axis of symmetry, and with the radius of Earth for radius. The part of the cylinder ``behind'' the Earth is in shadow. Thus if, on a particular day, LARES experiences an eclipse, then its orbit will intersect the cylinder in 4 points, 2 of which are $t_{entrance}$ and $t_{exit}$. To pick out the correct two solutions, we consider the dot product $\hat{r}_{Sun} \cdot \hat{r}_{sat}$. When this dot product is negative, the vector $\hat{r}_{Sun}$ makes an angle larger than $\frac{\pi}{2}$ with the vector $\hat{r}_{sat}$, and there is eclipse.

\section{Computation of the temperatures and the thermal drag}\label{Sec:Analysis}
In this section we will compute the temperature of each isothermal element, and subsequently the thermal drag.
\subsection{Surface condition of the CCRs}\label{surfacecond}
Appendix \ref{App:Constants} lists parameters for our model, including the absorptivity/emissivity of the fused silica CCRs both in the visible and in the IR. They are quoted both for clean glass - measured in lab, and for ``on-orbit'', or ``dirty'' glass. Absorptivity and emissivity are surface properties, and are modified by surface contamination. The on-orbit visual and IR values are important because they determine the heat balance in the CCRs.\\
To begin with the values in the visual, clean Suprasil glass has a very small absorptivity in the visual (a few parts per million \cite{Carpenteretal}), and is quite dark in the IR ($\epsilon_{IR} = \alpha_{IR} =0.82$). As Slabinski \cite{Slabinski(1996)} pointed out, an aside to discuss {\it Optical Solar Reflectors (OSRs)} is useful at this point. OSRs are silvered glass mirrors which are placed  (silver side in) on spacecraft exteriors to control temperature in satellites exposed to sunlight. The silver reflects most of the incident visual energy, but the glass, which has a high emissivity in the IR, efficiently radiates heat away.\\
In the 1970s it was observed that OSRs became less effective over time. At least two experiments \cite{PenceGrant(1981),Hyman} were flown with a carefully designed and modeled configuration of heater, thermocouples, and OSRs, to measure the visual absorptivity of the OSRs. It was found that the visual absorptivity increased from about 10 \% after launch to over 15 \% over a period of approximately a year. In his exhautive paper on thermal effects in the LAGEOS satellite, Slabinski \cite{Slabinski(1996)} took a value of $\alpha_{vis} = 0.15$. He quotes conversations with aerospace designers that ``surfaces in the space environment tend toward gray. Initially white surfaces turn to gray; initially black surfaces turn to gray'' (Slabinski, private communication 2015). We follow Slabinski by taking the on-orbit $\alpha_{vis}$ = 0.15 for the CCRs.\\
However, Slabinski uses a clean glass IR emissivity: 0.82. We show below that the somewhat smaller value of 0.60 matches the LARES observations more closely. Physically, any modification to the glass surface should change the IR properties as well as the visible properties, and moving the IR emissivity $0.82 \rightarrow 0.60$ moves in the direction of graying the dark IR surface of the clean glass. Thus we regard an on-orbit value of $0.60$ a plausible value of IR emissivity for the CCRs. Reducing $\epsilon_{IR} = \alpha_{IR}$ reduces the thermal response of the satellite to warming by the Earth, which reduces the thermal drag. We present results for both $\epsilon_{IR} = \alpha_{IR} = 0.82$, and for $\epsilon_{IR} = \alpha_{IR} = 0.60$, below.
\subsection{The radiation intensities}
\begin{figure}
$$
\begin{array}{cc}
 \includegraphics[width=7cm]{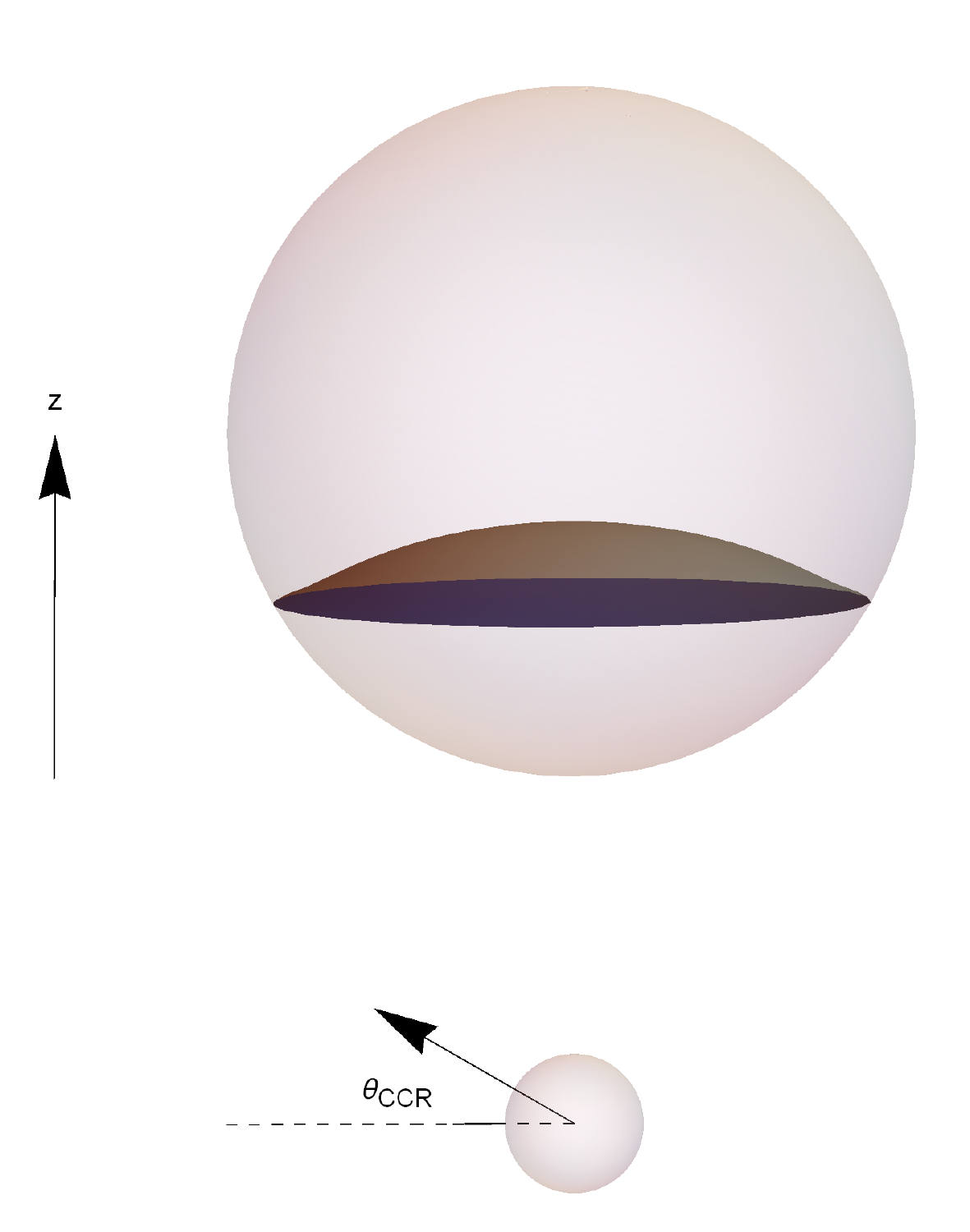}
\end{array}
$$
\caption{The setup used to compute the Earth's IR radiation on the satellite $I_{IR}(\theta_{CCR})$, with the larger sphere representing Earth and the smaller sphere representing LARES. Here we have depicted a situation with $\theta_{CCR} > 0$ (i.e. the CCR faces toward Earth). The portion of the celestial sphere subtended by the Earth from LARES, over which we integrate to obtain $I_{IR}$, is the spherical cap, which has an angular radius 54.55 degrees. The separation of LARES and the Earth is not to scale in the figure.}
\label{FinitesizeEarthfig}
\end{figure}
\begin{figure}
$$
\begin{array}{cc}
 \includegraphics[width=12cm]{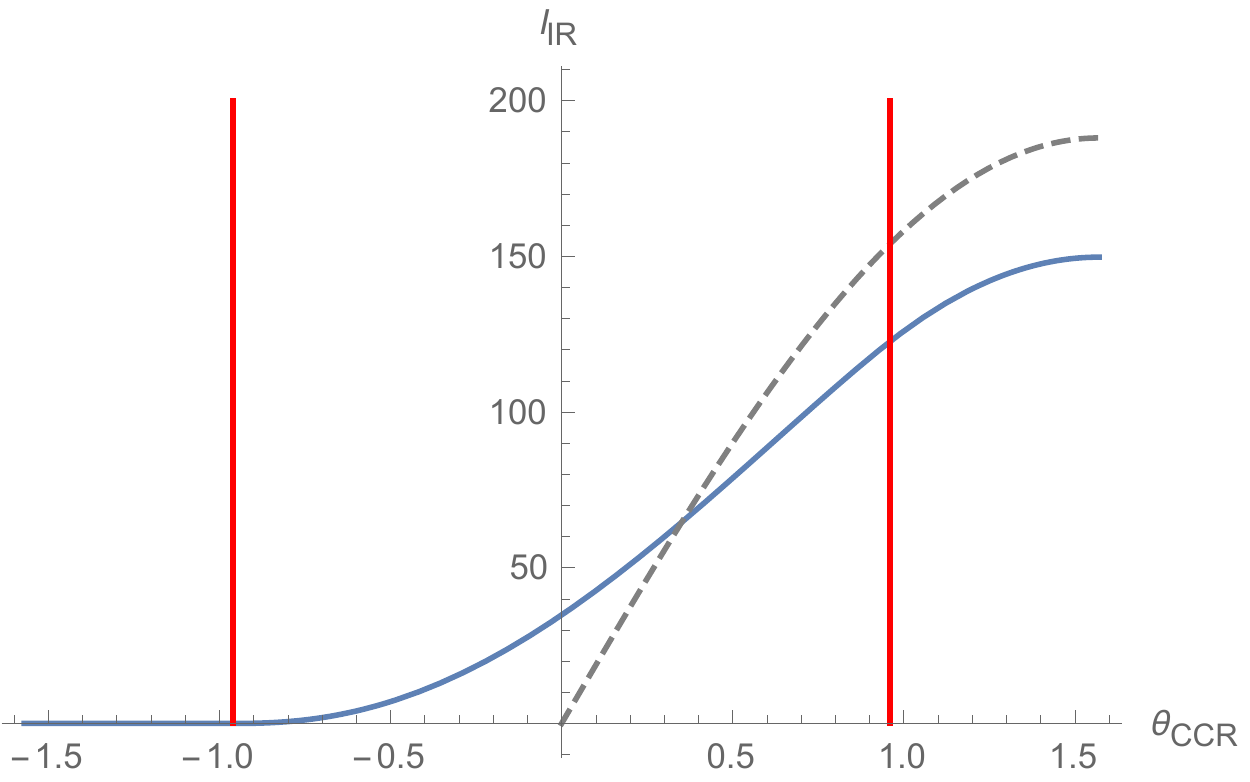}
\end{array}
$$
\caption{Plot of the function $I_{IR}{(\theta_{CCR})}$ in blue. The two vertical red lines correspond to $\theta_{CCR} = \pm \alpha_{e}$. For comparison, we have also included the intensity profile if the Earth was approximated as a point source (in dashed gray).}
\label{IIRfig}
\end{figure}

First we will need the radiation intensity incident on each CCR. The contribution from sunlight is quite straightforward. First we compute the angle between the normal direction to the CCR and the sunlight direction:
\begin{equation}
\cos{(\gamma_{vis}{(t,\theta_{I},k)})} = \hat{r}_{CCR}{(t,\theta_{I})} \cdot \hat{r}_{Sun}{(k)}
\end{equation}
Notice that we are assuming the Sun to be a point source, so that all the sunlight comes from the direction along the Earth-Sun axis, since the Sun subtends a very small solid angle in the sky, as seen from LARES. The visible intensity incident on the CCR is then:
\begin{equation}
I_{vis}{(t,\theta_{I},k)} = \Phi \cos{(\gamma_{vis}{(t,\theta_{I},k)})} \Theta{(\cos{(\gamma_{vis}{(t,\theta_{I},k)})})}
\end{equation}
where the unit step function $\Theta$ ensures that, when the CCR faces away from sunlight, the intensity vanishes (at any moment only half the satellite is illuminated by sunlight). Moreover the above is only valid outside the solar eclipse; the visible intensity also vanishes when the satellite is in the eclipse.\\
Next, consider the Earth IR radiation. Here we hit a substantial complication owing to the fact that the Earth, unlike the Sun, cannot be modelled as a point source. We have to take into account the finite-size effects of the Earth. Fortunately there is a remarkable simplification if we assume the Earth to be a Lambertian emitter: as viewed from a unit surface area of the satellite, the Earth's intensity per unit solid angle is a constant, given by $N_{IR} = 71 \mathrm{W} \cdot \mathrm{m}^{-2} \cdot \mathrm{sr}^{-1}$ (\cite{Slabinski(1996)}). More explicitly, this means that for each unit of solid angle of the source (i.e. the IR Earth) as seen from the satellite, a unit area of the receiver's surface receives 71 Watts if oriented perpendicularly to the radiation (if the orientation is not perpendicular, the power will be reduced by a Lambertian cosine factor, as usual). Thus, the total IR intensity is obtained by evaluating an integral of the solid angle subtended by Earth as viewed from LARES, weighted by the Lambertian cosine factor, then multiplying the result by $N_{IR}$.\\
To evaluate the integral described above, we set up spherical coordinate system $(r,\theta,\phi)$ centered on the satellite such that the z-axis is along the satellite-Earth axis (see fig. \ref{FinitesizeEarthfig}). Then the Earth subtends the disk $\theta \leq \alpha_{e}$ = 54.55 degrees, where
\begin{equation}
\alpha_{e} = \arcsin{\left(\frac{R_{E}}{a}\right)}
\end{equation}
where $R_{E}$ is the radius of Earth's IR-emitting atmosphere \footnote{\label{REfootnote} Technically, for the purpose of the eclipse, we should use the actual Earth radius (6378 km) rather than the radius of the IR-emitting atmosphere, which is slightly larger (6407 km). However the difference between the two radii introduces a negligible error.}. Moreover, notice that there is no special direction for the spin axis in these coordinates. Thus we can without loss of generality rotate coordinates around the z-axis to assume that the CCR momentarily lies in the x-z plane. Then the unit vector $\hat{r}_{CCR}$ from the satellite's center to the CCR can be parametrized by one angle $\theta_{CCR}$ (see fig. \ref{FinitesizeEarthfig}):
\begin{equation}
\hat{r}_{CCR} = (\cos{\theta_{CCR}},0,\sin{\theta_{CCR}})
\end{equation}
with $\theta_{CCR}$ ranging from $-\frac{\pi}{2}$ to $\frac{\pi}{2}$. Then we can write down the cosine of the angle between $\hat{r}_{CCR}$ and the vector from the CCR to some area element on the Earth disk with coordinate $(\theta,\phi)$:
\begin{equation}\label{cosgammaIR}
\cos{(\gamma_{IR}{(\theta_{CCR},\theta,\phi)})} = \cos{\theta_{CCR}}\sin{\theta}\cos{\phi} + \sin{\theta_{CCR}}\cos{\theta}
\end{equation}
The IR intensity on the CCR is now only a function of $\theta_{CCR}$. It can be piecewise defined as follows. If $\theta_{CCR} < -\alpha_{e}$, then by inspection of fig. \ref{FinitesizeEarthfig}, we can easily seen that the CCR does not intercept any IR radiation. In this case $I_{IR}{(\theta_{CCR})} = 0$. If $-\alpha_{e} < \theta_{CCR} < 0$, the CCR sees less than half the Earth disk, and
\begin{equation}
I_{IR}{(\theta_{CCR})} = 2N_{IR} \int_{|\theta_{CCR}|}^{\alpha_{e}} \sin{\theta}d\theta \int_{0}^{F{(\theta_{CCR},\theta)}} \cos{(\gamma_{IR}{(\theta_{CCR},\theta,\phi)})} d\phi
\end{equation}
with $F{(\theta_{CCR},\theta)}$ obtained by setting equation (\ref{cosgammaIR}) to zero and solving for $\phi$:
\begin{equation}
F{(\theta_{CCR},\theta)} = \arccos{\left(-\frac{\tan{\theta_{CCR}}}{\tan{\theta}}\right)}
\end{equation}
Next, if $0 < \theta_{CCR} < \alpha_{e}$, then the CCR sees more than half of the Earth disk, and
\begin{eqnarray}
I_{IR}{(\theta_{CCR})} &=& 2N_{IR} \int_{|\theta_{CCR}|}^{\alpha_{e}} \sin{\theta}d\theta \int_{0}^{F{(\theta_{CCR},\theta)}} \cos{(\gamma_{IR}{(\theta_{CCR},\theta,\phi)})} d\phi \nonumber \\
&+& 2\int_{0}^{|\theta_{CCR}|}\sin{\theta}d\theta \int_{0}^{\pi} \cos{\gamma_{IR}{(\theta_{CCR},\theta,\phi)}}d\phi 
\end{eqnarray}
Finally, when $\alpha_{e} < \theta_{CCR}$, then the CCR views the whole Earth disk, and the integral can be evaluated analytically:
\begin{equation}
I_{IR}{(\theta_{CCR})} = \pi N_{IR} \sin{\theta_{CCR}} \sin^{2}{\alpha_{e}}
\end{equation}
The function $I_{IR}{(\theta_{CCR})}$ is plotted in fig. \ref{IIRfig}. Finally, $\theta_{CCR}$ itself is a function of the time $t$ elapsed during one orbit, the number of days $k$ since launch, and the colatitude $\theta_{I}$ of the CCR:
\begin{equation}
\theta_{CCR}{(t,k,\theta_{I})} = -\arcsin{\left(\hat{r}_{CCR}{(t,\theta_{I})} \cdot \hat{r}_{sat}{(t,k)}\right)}
\end{equation}
We close this subsection by a few remarks on the error that would be introduced if we were modelling the Earth as a point source. One can multiply $N_{IR}$ by the total solid angle $\Delta\Omega$ subtended by Earth as viewed from LARES to get an effective power per unit receiver area $W_{IR}$ (of the dimension $\mathrm{W \cdot m^{-2}}$). The solid angle $\Delta \Omega$ is easily computed:
\begin{equation}
\Delta\Omega = 2\pi \int_{0}^{\alpha_{e}}\sin{\theta}d\theta = 2\pi\left(1-\cos{\alpha_{e}}\right)
\end{equation}
We then find $W_{IR} = 188 \mathrm{W \cdot m^{-2}}$. We must multiply this flux by the cosine of the angle between the CCR normal and the direction of the radiation, i.e. $\cos{\left(\frac{\pi}{2}-\theta_{CCR}\right)}$ to obtain the power absorbed by a surface element. In fig. \ref{IIRfig}, we also plot this resulting intensity profile for comparison to the correct treatment. As can be seen from the plot, the finite-size effect of the Earth spreads out the intensity more evenly across the satellite surface, thus significantly reducing the thermal drag.

\subsection{The equations of conservation and the temperatures}
In this subsection, we write down the equations of conservation of energy for each CCR and the metal, Fourier transform those equations and solve for their temperatures. For a CCR of colatitude $\theta_{I}$, conservation of energy reads:
\begin{equation}
\epsilon_{eff}A_{gl}\sigma(T_{W}^{4}{(t)}-T_{I}^{4}{(t)}) + \epsilon_{gl,IR}\pi R^{2} I_{IR}{(\theta_{I},t)} + \alpha_{gl,vis}\pi R^{2} I_{vis}{(\theta_{I},t)} - \epsilon_{gl,IR}\sigma \pi R^{2} T_{I}^{4}{(t)} = m_{CCR} c_{gl}\frac{dT_{I}{(t)}}{dt}
\end{equation}
where  $T_{I}$ is the CCR temperature, which only depends on the row $I$ since we are in the fast-spin regime. The terms on the left-hand side represent the following contributions: the first term is the net heat tranferred to the CCR from the cavity, the second term is the heat absorbed at the exterior surface from Earth IR radiation, the third term is the heat absorbed at the exterior surface from sunlight, and the fourth term is the heat lost into space by radiation. On the right hand side is the rate of change of the temperature with time.\\
For the metal, conservation of energy reads:
\begin{equation}\label{conservationW}
P_{W,vis}{(t)} + P_{W,IR} - A_{gl}\epsilon_{eff}\sigma\left(92T_{W}^{4}{(t)}-\sum_{I} n_{I} T_{I}^{4}{(t)}\right) - \epsilon_{W,IR}A_{W,vac}\sigma T_{W}^{4}{(t)} = m_{W}c_{W}\frac{dT_{W}{(t)}}{dt}
\end{equation}
The first term on the left is the power absorbed by the metal from sunlight. Its time-dependence is solely due to the eclipse. When the satellite is outside the eclipse, it is approximately constant and approximately given by:
\begin{equation}
P_{W,vis} = \alpha_{W,vis}\pi R_{sat}^{2}\Phi - \alpha_{W,vis}\Phi \pi R^{2} + \left[\frac{1}{2}(1-\alpha_{gl,vis})-\alpha_{gl,vis}\right]\pi R^{2}\Phi \sum_{I \neq 1} n_{I}\cos{\theta_{I}}
\end{equation}
The power absorbed by the metal from sunlight is in fact a funtion of the orientation of the satellite, and hence of time, but we assume that it is approximately constant, and evaluate for a specific configuration, namely when one pole of the satellite is directly pointed at the Sun. Then the first term above is the power absorbed as if the satellite were a simple metal sphere. The second term above is the power incident on the North Pole CCR which, instead of being absorbed, is reflected back. The last term above accounts for the remaining CCRs (other than the North Pole). We assume that a fraction $\frac{1}{2}(1-\alpha_{gl,vis})$ of sunlight incident on these CCRs make it into the cavity and ends up absorbed by the metal (see \cite{Slabinski(1996)}).\\
The second term in (\ref{conservationW}) represents the power absorbed by the metal from the IR. It is approximately constant in time:
\begin{equation}
P_{W,IR} = \epsilon_{W,IR}R_{sat}^{2}2\pi \int_{0}^{\frac{\pi}{2}+\alpha_{e}} \sin{\theta} I_{IR}{\left(\frac{\pi}{2}-\theta\right)}d\theta - \epsilon_{W,IR}\pi R^{2} \sum_{I} n_{I} I_{IR}{\left(\frac{\pi}{2}-\theta_{I}\right)}
\end{equation}
where the first term on the right-hand side represents the power absorbed as if the satellite were a metal sphere, and the remaining terms account for the CCRs. Coming back to equation (\ref{conservationW}), the third term on the left-hand side represents the power transferred to the CCRs from the metal, and the 4th term on the left-hand side is the power radiated into space.\\
Since we expect the time variation in the temperature to be small, we linearize the conservation equations by letting
\begin{equation}
T_{W}{(t)} = T_{W,0} + \Delta T_{W}{(t)}
\end{equation}
\begin{equation}
T_{I}{(t)} = T_{I,0} + \Delta T_{I}{(t)}
\end{equation}
with $\Delta T_{W} << T_{W,0}$ and $\Delta T_{I} << T_{I,0}$. Then:
\begin{equation}
T_{W}^{4}{(t)} \approx T_{W,0}^{4} + 4T_{W,0}^{3}\Delta T_{W}{(t)}
\end{equation}
\begin{equation}
T_{I}^{4}{(t)} \approx T_{I,0}^{4} + 4T_{I,0}^{3}\Delta T_{I}{(t)}
\end{equation}
Moreover, thanks to the periodicity of the motion, we can expand the time-dependent pieces of the temperatures into Fourier series \footnote{Notice that we carry out a Fourier sum, not a Fourier integral. To do this we are implicitly assuming that the spin frequency $\omega_{s}$ is an integer multiple of the orbital frequency $\omega_{o}$, so that all time-dependent quantities are exactly periodic. In the rapid-spin regime, this does not cause any difficulty (in particular, no resonances).}:
\begin{equation}
\Delta T_{W}{(t)} = \sum_{n \in \mathcal{Z}, n \neq 0} T_{W,n} e^{i\omega_{0} n t}
\end{equation}
\begin{equation}
\Delta T_{I}{(t)} = \sum_{n \in \mathcal{Z}, n \neq 0} T_{I,n} e^{i\omega_{0} n t}
\end{equation}
Decomposing the conservation equations into modes, we find, for the time-independent mode:
\begin{equation}
\epsilon_{eff} A_{gl}\sigma (T_{W,0}^{4}-T_{I,0}^{4}) + \epsilon_{gl,IR}\pi R^{2}I_{IR,0}{(\theta_{I})} + \epsilon_{gl,vis}\pi R^{2}I_{vis,0}{(\theta_{I})} - \epsilon_{gl,IR}\sigma \pi R^{2} T_{I,0}^{4} = 0
\end{equation}
\begin{equation}
P_{W,vis,0} + P_{W,IR} - A_{gl}\epsilon_{eff}\sigma (92T_{W,0}^{4} - \sum_{I}n_{I}T_{I,0}^{4}) - \epsilon_{W,IR}A_{W,vac}\sigma T_{W,0}^{4} = 0
\end{equation}
where $P_{W,vis,0}$, $I_{IR,0}$ and $I_{vis,0}$ are the time-averaged $P_{W,vis}$, $I_{IR}$ and $I_{vis}$, respectively. All of these three quantities can be computed by numerical integration. For the $n=1$ mode, we have:
\begin{equation}
4\epsilon_{eff}A_{gl}\sigma(T_{W,0}^{3}T_{W,1}-T_{I,0}^{3}T_{I,1}) + \epsilon_{gl,IR}\pi R^{2} I_{IR,1}{(\theta_{I})} + \alpha_{gl,vis}\pi R^{2}I_{vis,1}{(\theta_{I})} - 4\epsilon_{gl,IR}\sigma \pi R^{2} T_{I,0}^{3}T_{I,1} = m_{CCR}c_{gl}i\omega_{o} T_{I,1}
\end{equation}
\begin{equation}
P_{W,vis,1} - 4A_{gl}\epsilon_{eff}\sigma(92T_{W,0}^{3}T_{W,1}-\sum_{I}n_{I}T_{I,0}^{3}T_{I,1}) - 4\epsilon_{W,IR}A_{W,vac}\sigma T_{W,0}^{3}T_{W,1} = m_{W}c_{W}i\omega_{o} T_{W,1}
\end{equation}
For the $n=2$ mode, we have:
\begin{equation}
4\epsilon_{eff}A_{gl}\sigma(T_{W,0}^{3}T_{W,2}-T_{I,0}^{3}T_{I,2}) + \epsilon_{gl,IR}\pi R^{2} I_{IR,2}{(\theta_{I})} + \alpha_{gl,vis}\pi R^{2}I_{vis,2}{(\theta_{I})} - 4\epsilon_{gl,IR}\sigma \pi R^{2} T_{I,0}^{3}T_{I,2} = 2m_{CCR}c_{gl}i\omega_{o} T_{I,2}
\end{equation}
\begin{equation}
P_{W,vis,2} - 4A_{gl}\epsilon_{eff}\sigma(92T_{W,0}^{3}T_{W,2}-\sum_{I}n_{I}T_{I,0}^{3}T_{I,2}) - 4\epsilon_{W,IR}A_{W,vac}\sigma T_{W,0}^{3}T_{W,2} = 2m_{W}c_{W}i\omega_{o} T_{W,2}
\end{equation}
By solving the equations above, we obtain the coefficients $T_{W,0}$, $T_{W,1}$, $T_{W,2}$ and $T_{I,0}$, $T_{I,1}$, $T_{I,2}$. The temperatures are then:
\begin{equation}
T_{W}{(t)} = T_{W,0} + 2\mathrm{Re}{(T_{W,1})}\cos{\omega_{0} t} - 2\mathrm{Im}{(T_{W,1})}\sin{\omega_{0} t} + 2\mathrm{Re}{(T_{W,2})}\cos{2\omega_{0} t} - 2\mathrm{Im}{(T_{W,2})}\sin{2\omega_{0} t}
\end{equation}
\begin{equation}
T_{I}{(t)} = T_{I,0} + 2\mathrm{Re}{(T_{I,1})}\cos{\omega_{0} t} - 2\mathrm{Im}{(T_{I,1})}\sin{\omega_{0} t} + 2\mathrm{Re}{(T_{I,2})}\cos{2\omega_{0} t} - 2\mathrm{Im}{(T_{I,2})}\sin{2\omega_{0} t}
\end{equation}
\subsection{The thermal drag}
We assume the surface of the satellite to be Lambertian: the angular dependence of the intensity emitted by any surface element $I(\theta)$ is proportional to the cosine of the angle between the normal vector to the surface element and the observer's line of sight:
\begin{equation}
I(\theta)=I_{0} \cos{\theta}
\end{equation}
where $I_{0}$ is the intensity emitted in the normal direction. Integrating over all angles yields the total radiated intensity by the surface element, given by the blackbody radiation formula:
\begin{equation}
\epsilon \sigma T^{4} = \int I_{0} \cos{\theta} d\Omega = \pi I_{0}
\end{equation}
where $\epsilon$ is the emissivity of the surface element, $\sigma$ is the Stefan-Boltzmann constant, and $T$ is the temperature of the surface element. The total momentum emitted in a unit time by the surface element into an angular ring $d\Omega$ is:
\begin{equation}
dP(\theta) = \frac{I(\theta)}{c} d\Omega
\end{equation}
The force on the surface element is obtained by integrating over the hemisphere:
\begin{equation}\label{Forceonareaelement}
F = -\int \cos{\theta} dP(\theta) = -\frac{2}{3} \frac{\epsilon \sigma T^{4}}{c}
\end{equation}
where, by symmetry, it is enough to integrate over the normal component of $dP$. We emphasize that this expression, in particular the temperature $T$ appearing in equation \ref{Forceonareaelement}, refers to one infinitesimal surface ring of the satellite; the temperature varies across the satellite. Expression \ref{Forceonareaelement} must be integrated appropriately across the surface to determine the net thermal thrust.\\
In fact, the thermal thrust due to the metal is negligible because the tungsten is essentially isothermal (see equation (\ref{DeltaTlinearized})) and therefore radiates isotropically. The axial force is therefore supplied by a sum over CCRs:
\begin{equation}
F_{axial} = -\frac{2\epsilon_{gl,IR}\sigma\pi R^{2}}{3c} \sum_{I} n_{I}\cos{\theta_{I}}T_{I}^{4} 
\end{equation}
where the factor $\cos{\theta_{I}}$ serves to project on the spin axis.
\begin{figure}
$$
\begin{array}{cc}
  \includegraphics[width=9cm]{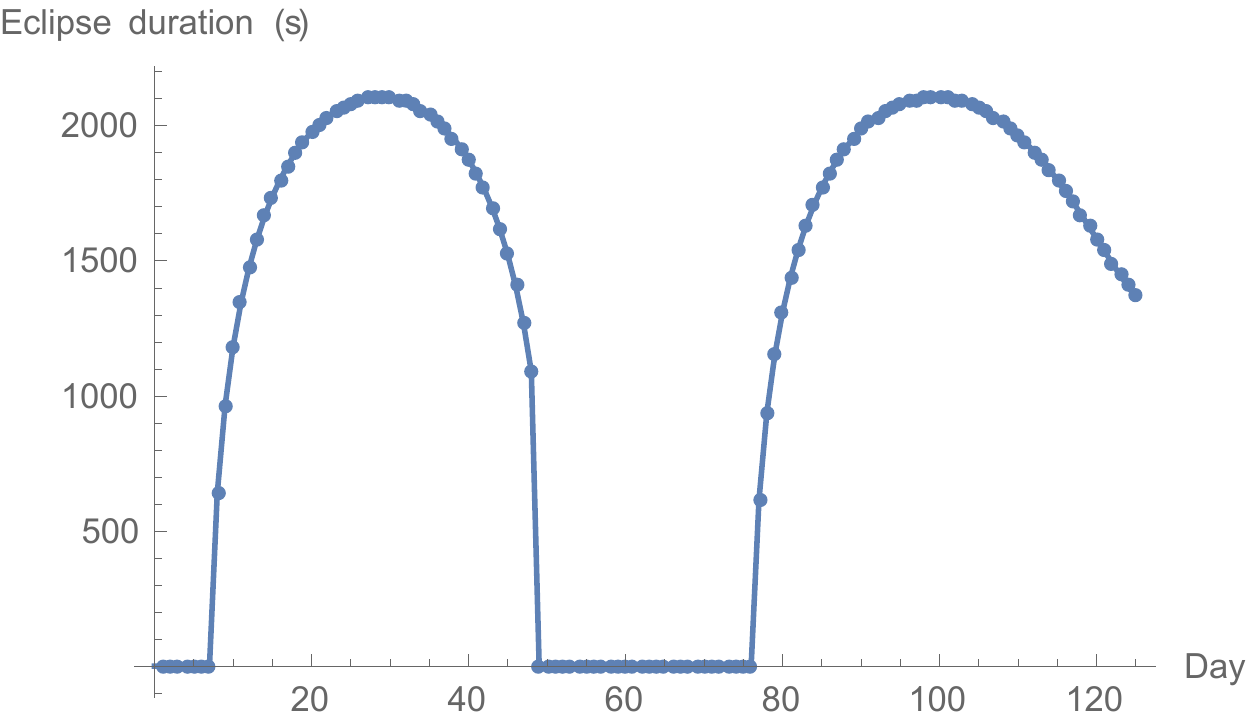} & \includegraphics[width=9cm]{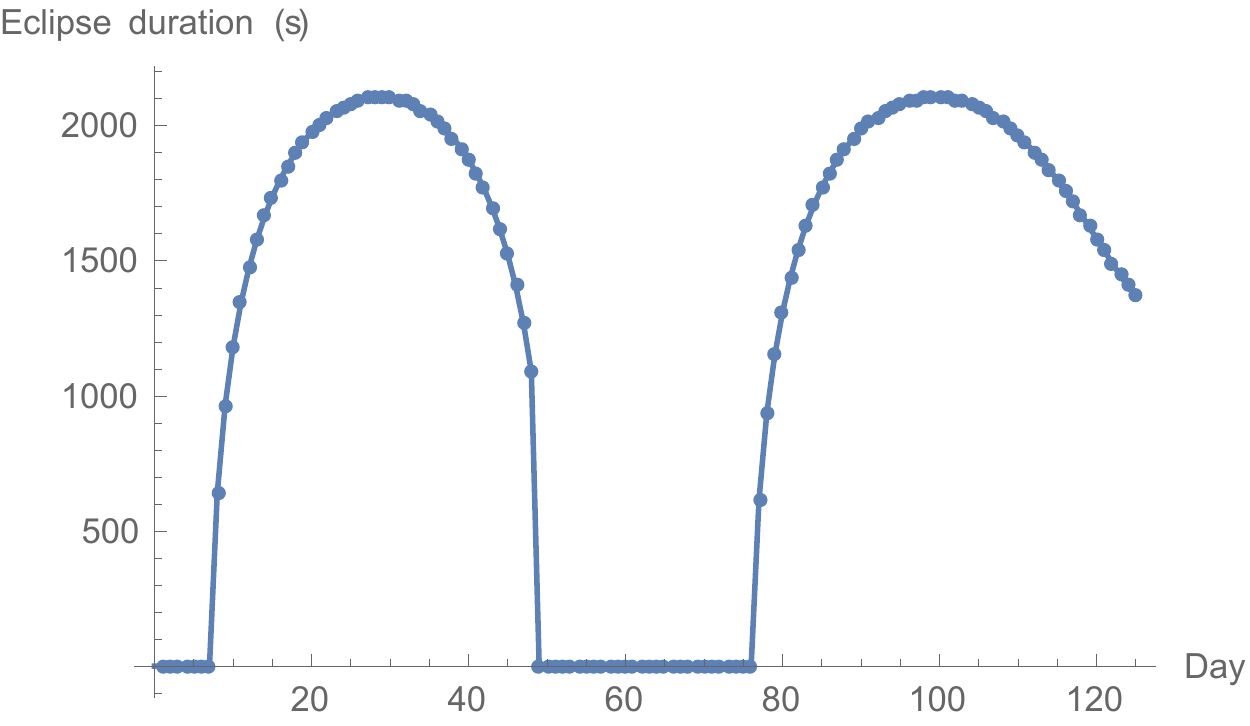} \\
  \includegraphics[width=9cm]{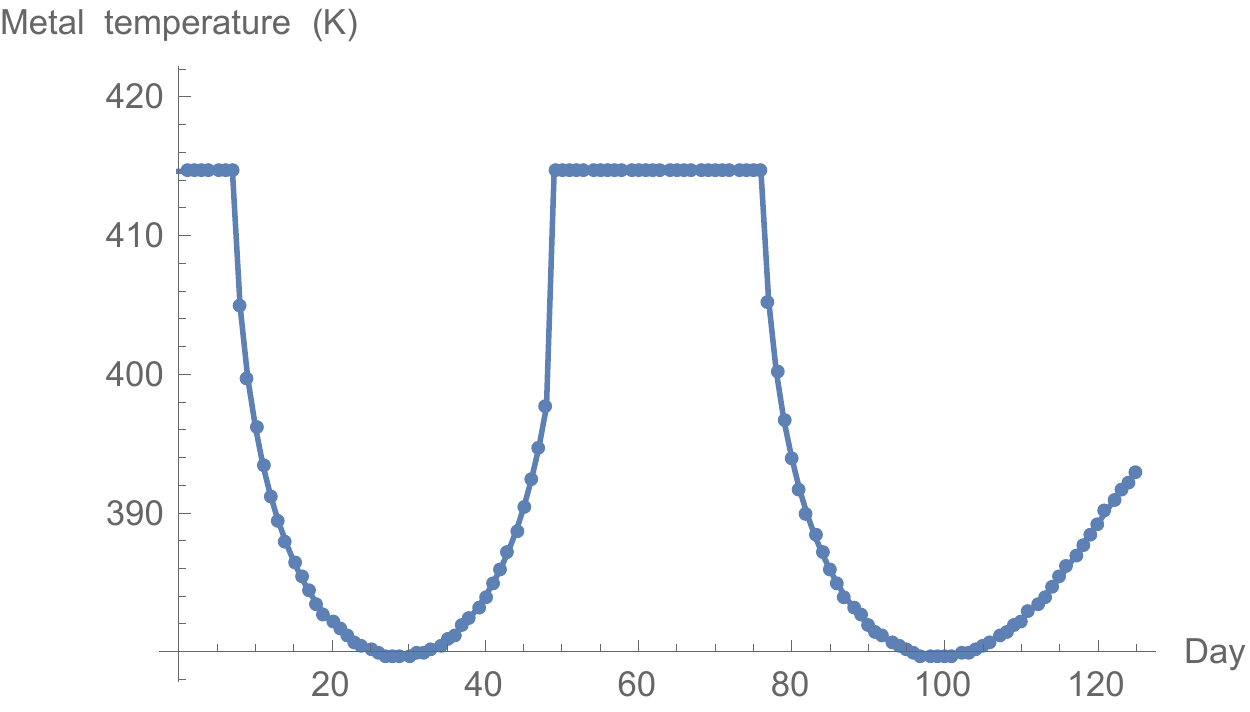} & \includegraphics[width=9cm]{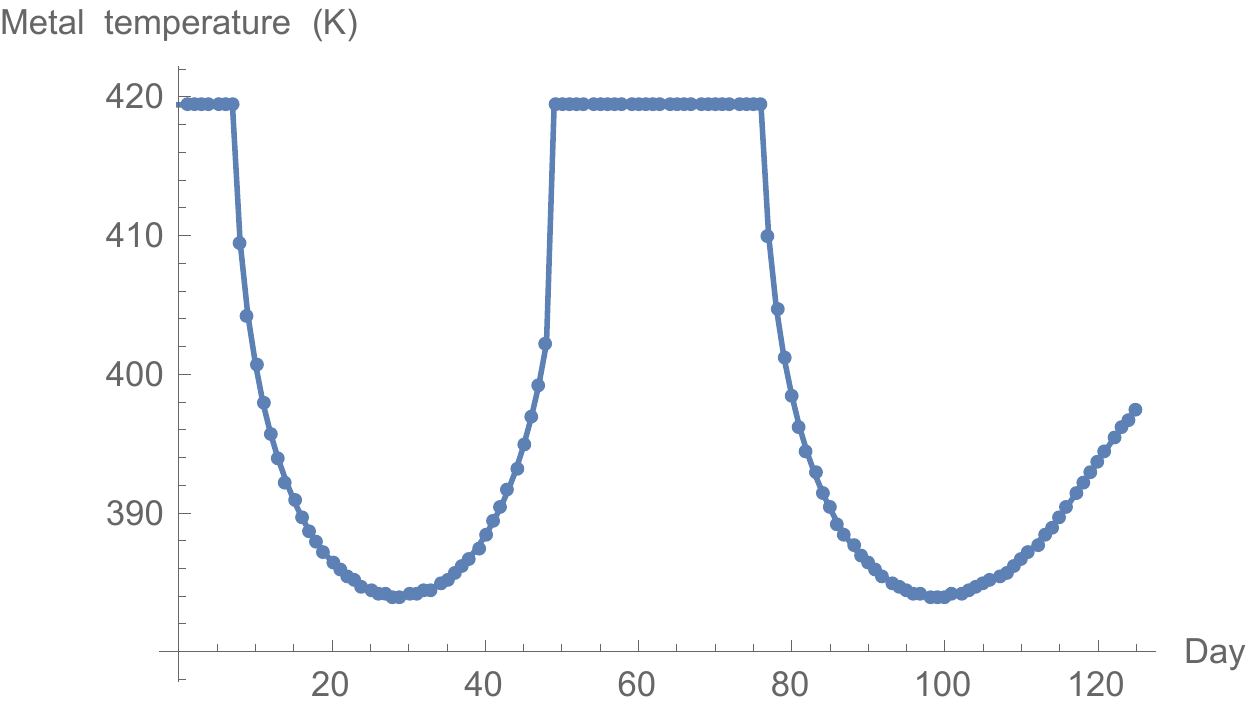}\\
  \includegraphics[width=9cm]{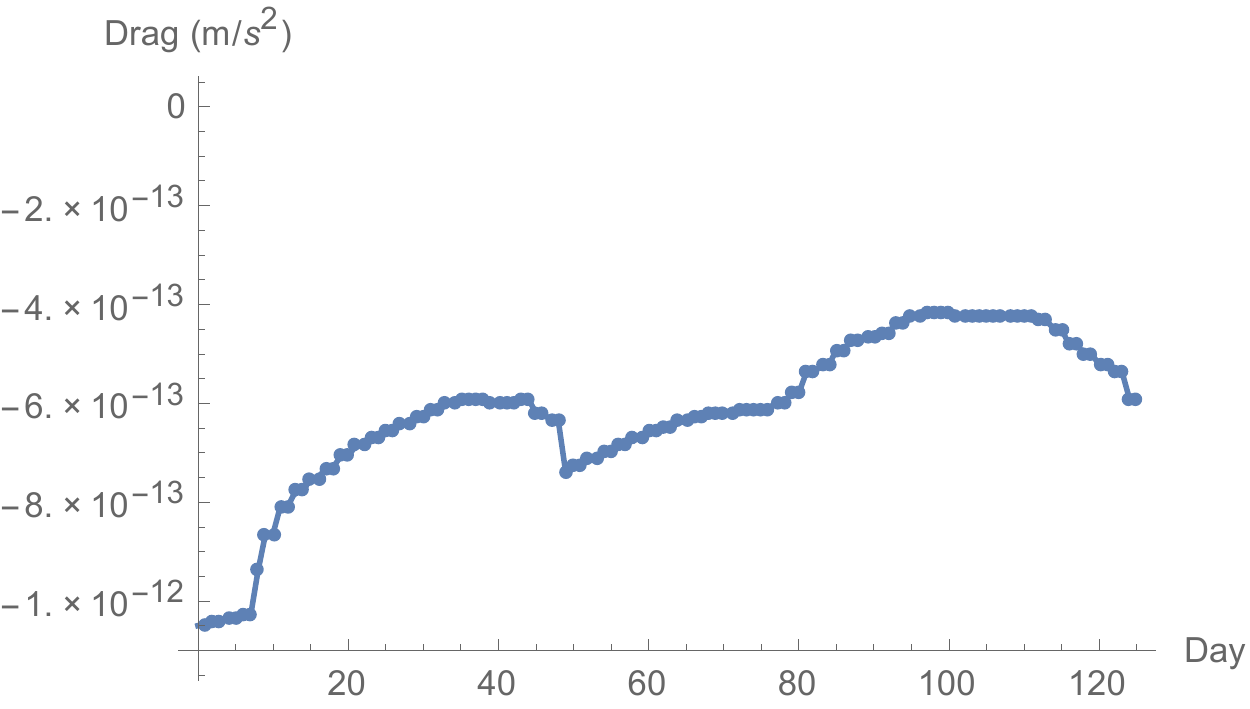} & \includegraphics[width=9cm]{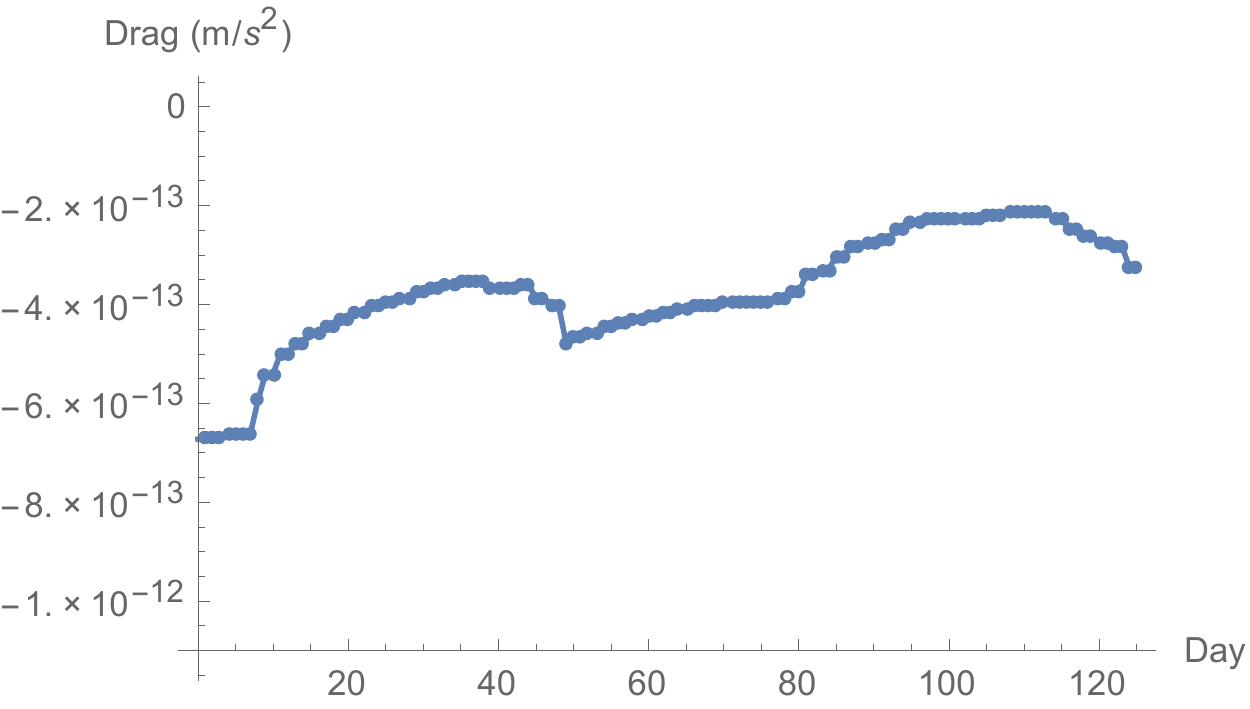}
\end{array}
$$
\caption{Histogram of the eclipse duration (top panels), the daily average metal temperature (middle panels), and the thermal drag (bottom panels) as a function of number of days since launch, for 126 days, assuming the clean-glass IR absorptivity $\alpha_{gl,IR}=0.82$ (left column) and the dirty-glass IR absorptivity $\alpha_{gl,IR}'=0.6$. The average thermal drag over the last 120 days is $-0.59 \ \mathrm{pm/s}^{2}$ for clean glass and $-0.36 \ \mathrm{pm/s}^{2}$ for dirty glass.}
\label{126daysfig}
\end{figure}
Next, to obtain the along-track component, we project the axial force on the direction of motion. Then we average over one orbit, and divide by the satellite mass to obtain the along-track acceleration:
\begin{equation}
a_{along-track} = \frac{1}{m_{sat}T}\int_{0}^{T} F_{axial}{(t)} \hat{S} \cdot \hat{v}_{sat}{(t,k)} dt
\end{equation}
where $\hat{v}{(t,k)}$ is the normalized velocity vector of the satellite. On the left column of fig. \ref{126daysfig}, we show histograms of the thermal drag, the metal temperature and the eclipse duration for the first 126 days using the clean-glass absorptivity in the IR $\alpha_{gl,IR} = 0.82$. The average thermal drag over the last 120 days of the first 126-day period is found to be:
\begin{equation}
\left<a_{along-track}\right>_{120 \ \mathrm{days}} = -0.59 \ \mathrm{pm}/\mathrm{s}^{2}
\end{equation}
This is around 50 \% larger than the observed value of $-0.4 \mathrm{pm}/\mathrm{s}^{2}$. This discrepancy can probably be explained by the fact that the values of all the absorptivities/emissivities are not known accurately. To illustrate this, if we lower the absorptivity of glass in the IR to the value $\alpha_{gl,IR}'=0.6$, a still plausible ``dirty glass'' value, then the average along-track acceleration over the 120-day period is now:
\begin{equation}
\left<a_{along-track}\right>_{120 \ \mathrm{days}} = -0.36 \ \mathrm{pm}/\mathrm{s}^{2}
\end{equation}
This time we obtain a value slightly smaller than the observed one. This suggests that there are possibly small unmodelled effects contributing to the observed drag $-0.4 /\mathrm{pm/s}^{2}$. The eclipse duration, metal temperature and thermal drag for this ``dirty glass'' case are plotted on the right column in fig. \ref{126daysfig}. Also, figures \ref{TDay0}, \ref{TDay30}, \ref{TDay60} and \ref{TDay90} give the tungsten temperature and the temperatures of the various CCR rows, for days 0, 30, 60, and 90 of predictions, for both the clean glass $\alpha_{gl,IR} = 0.82$, and the dirty glass $\alpha_{gl,IR}' = 0.60$ cases. To facilitate comparison between these figures, the range of the plots on the left panels (for the metal temperature) is always 2 Kelvins, and the range of the plots on the right panels (for the CCR temperatures) is always 22 Kelvins.
\begin{figure}
$$
\begin{array}{cc}
  \includegraphics[width=8cm]{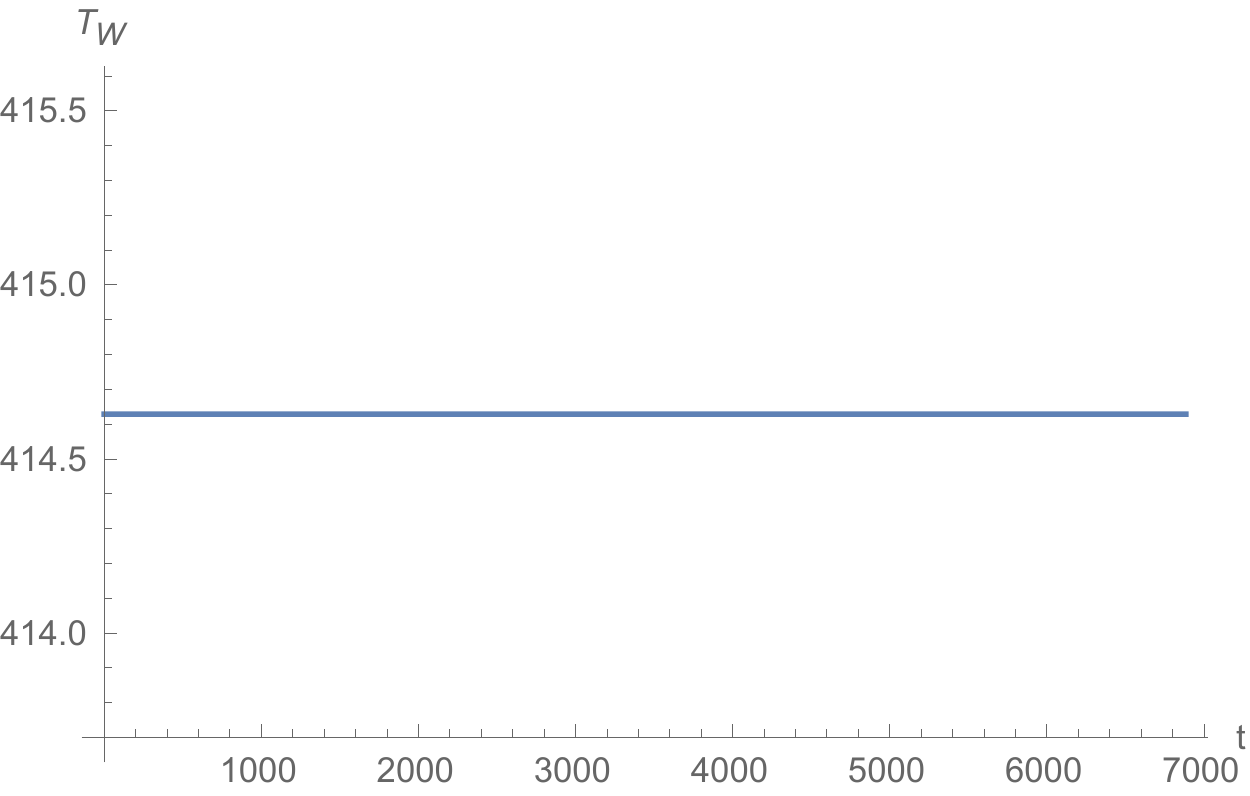} & \includegraphics[width=9cm]{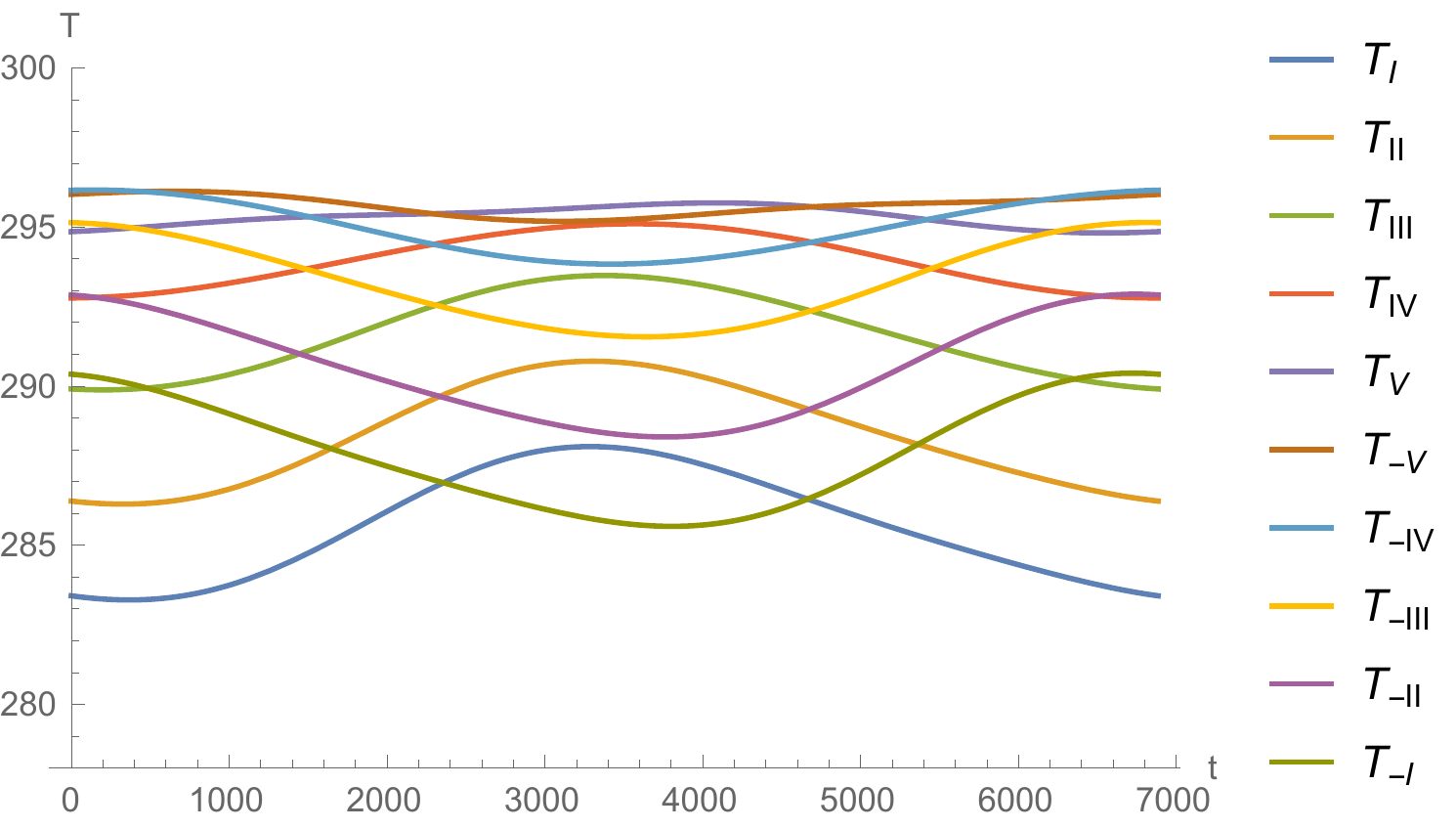} \\
  \includegraphics[width=8cm]{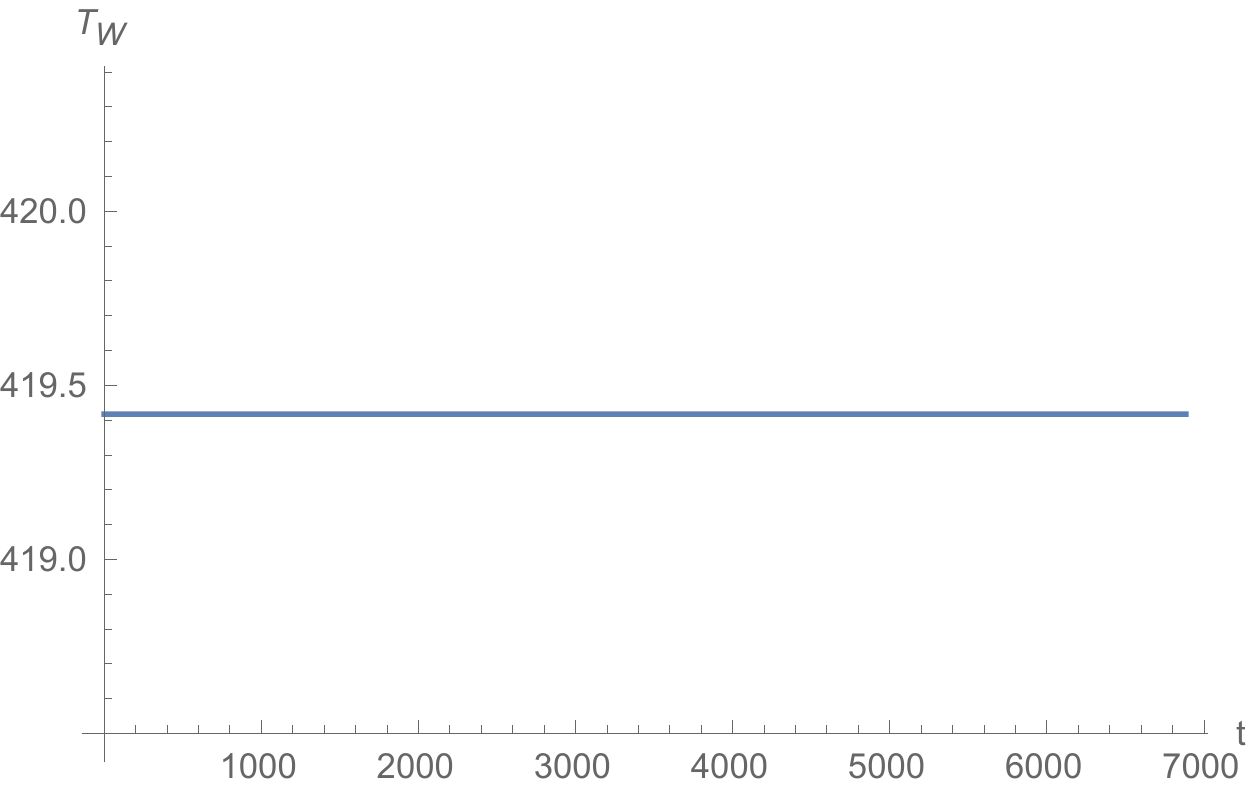} & \includegraphics[width=9cm]{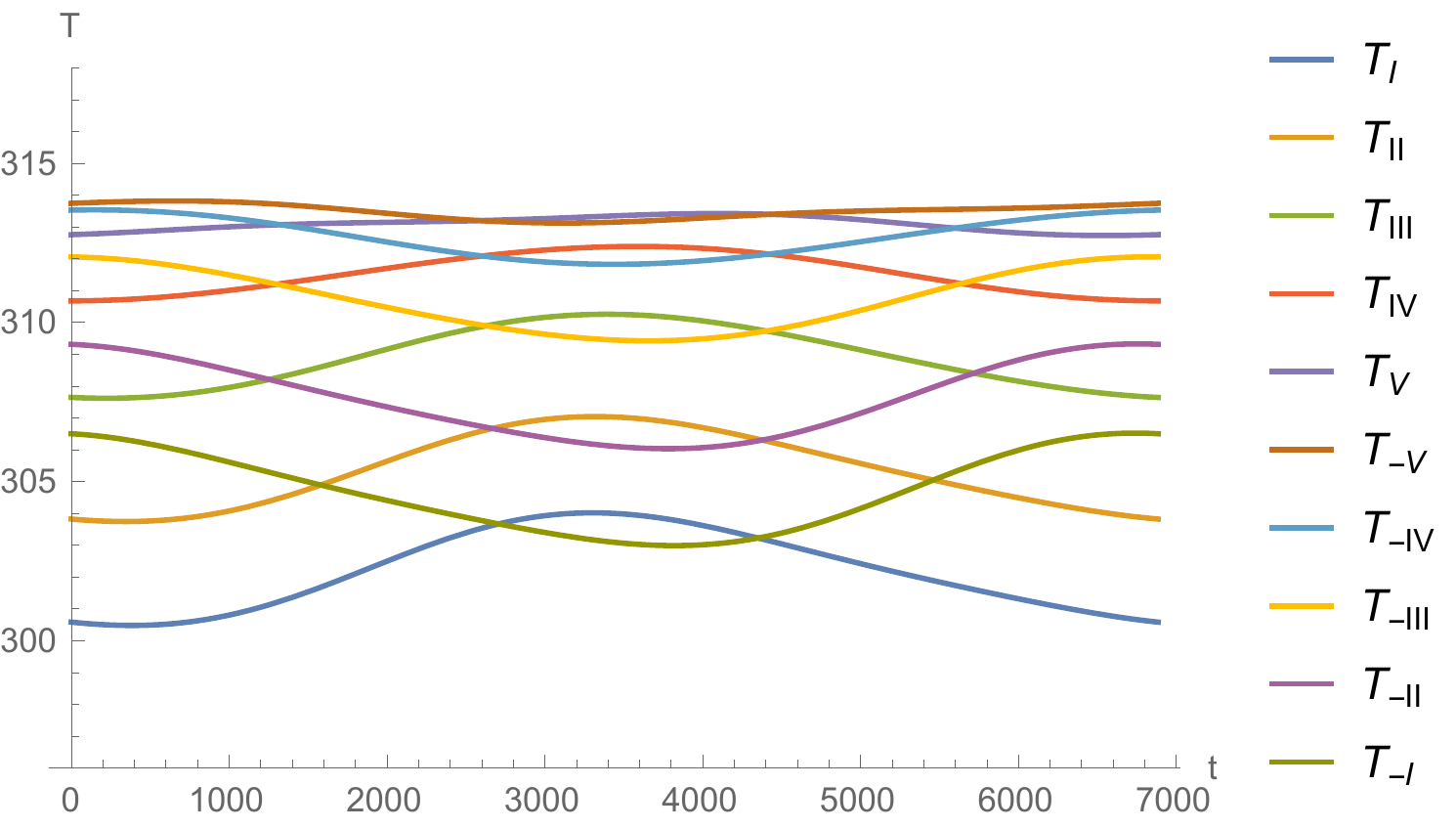}
\end{array}
$$
\caption{Top row: Plot of $T_{W}$ (left panel) and $T_{CCR}$ (right panel) (sum of the constant, first and second harmonics of the orbital frequency) over one period, on day 0, using the clean-glass absorptivity $\alpha_{gl,IR}=0.82$. There is no eclipse on this day, and as a result, the metal temperature is practically constant. The average along-track acceleration on this day is $-1.0 \ \mathrm{pm/s}^{2}$. For comparison, the average thermal drag over the 120-day period is $-0.59 \ \mathrm{pm/s}^{2}$. This day is not included in the average. Bottom row: Plot of $T_{W}$ (left panel) and $T_{CCR}$ (right panel) (sum of the constant, first and second harmonics of the orbital frequency) over one period, on day 0, using the dirty-glass absorptivity $\alpha_{gl,IR}'=0.6$. The average along-track acceleration on this day is $-0.67 \ \mathrm{pm/s}^{2}$. For comparison, the average thermal drag over the 120-day period is $-0.36 \ \mathrm{pm/s}^{2}$.}
\label{TDay0}
\end{figure}
\begin{figure}
$$
\begin{array}{cc}
  \includegraphics[width=8cm]{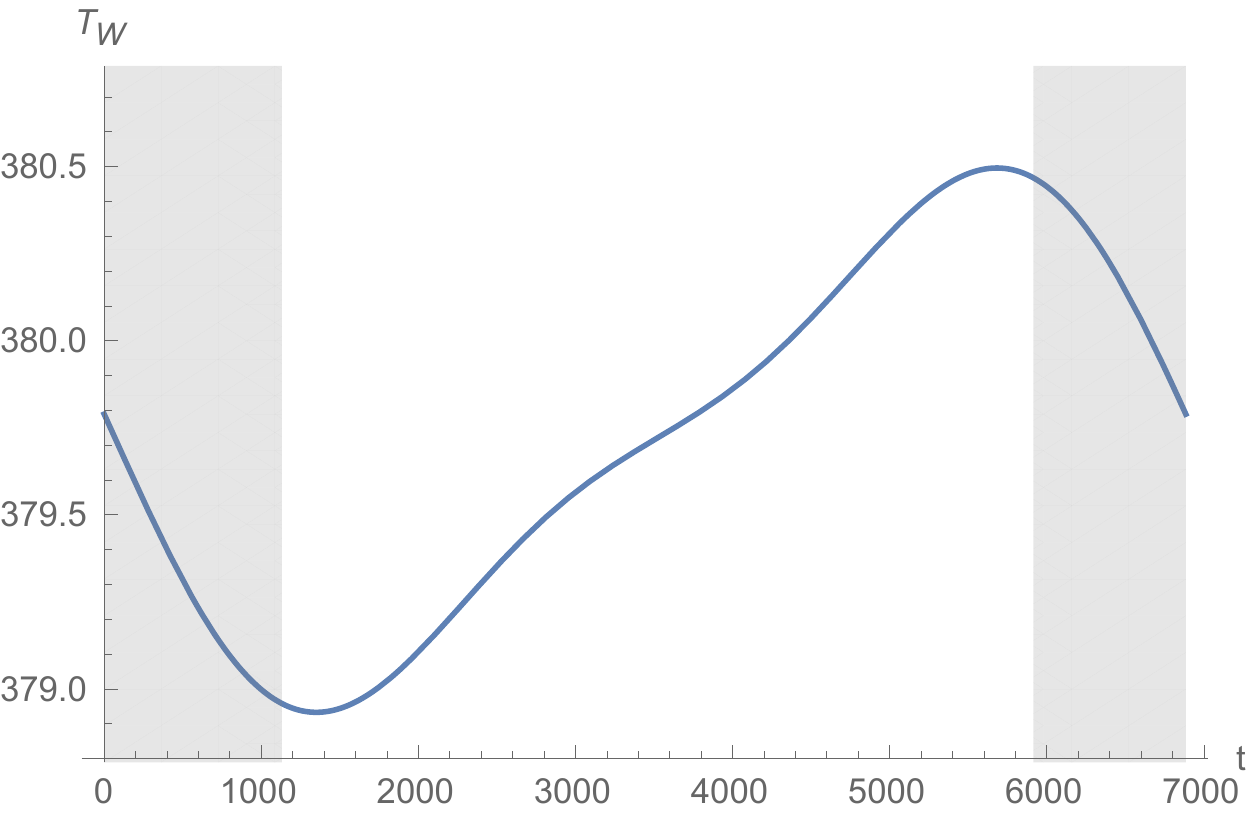} & \includegraphics[width=9cm]{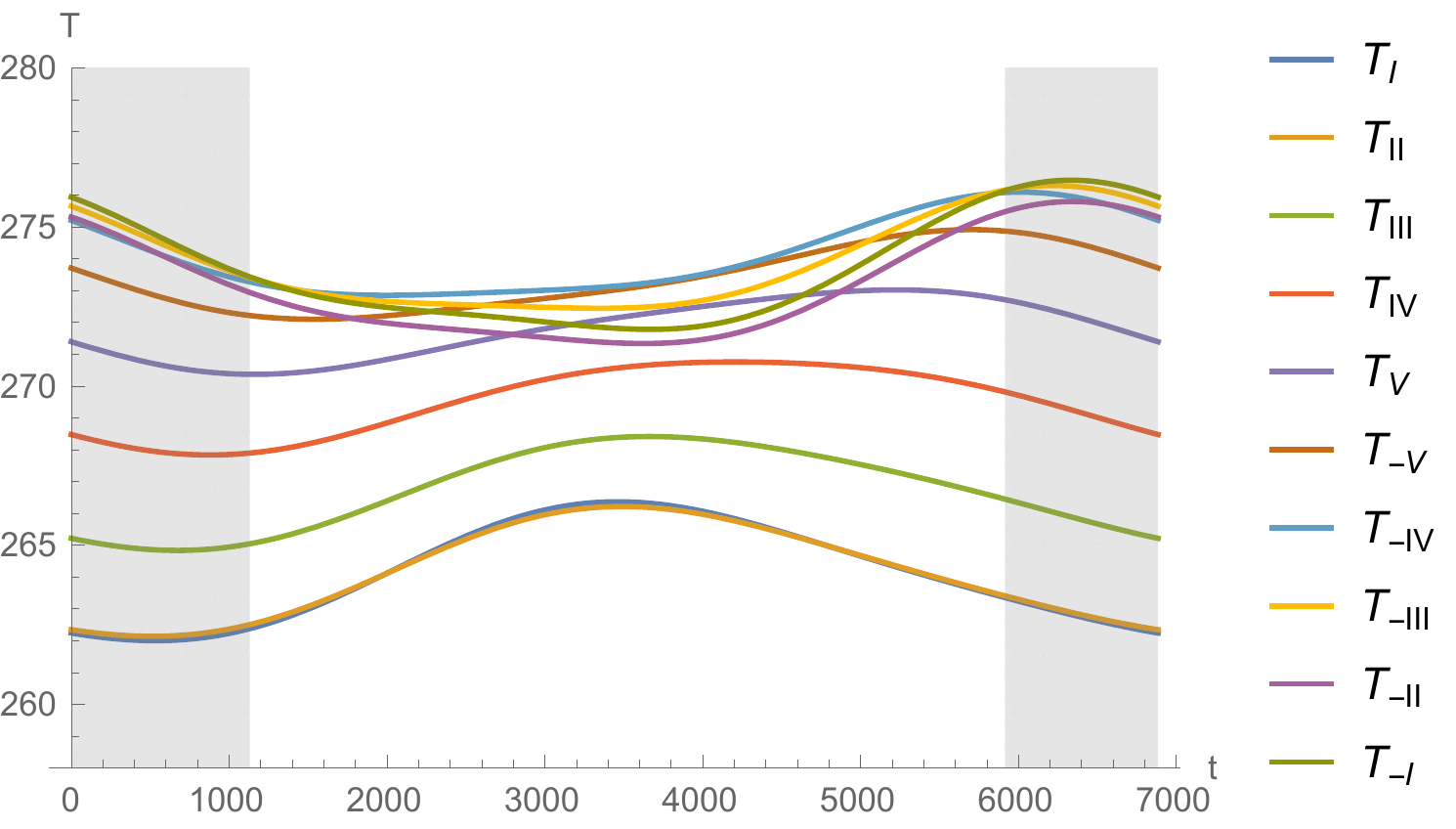} \\
  \includegraphics[width=8cm]{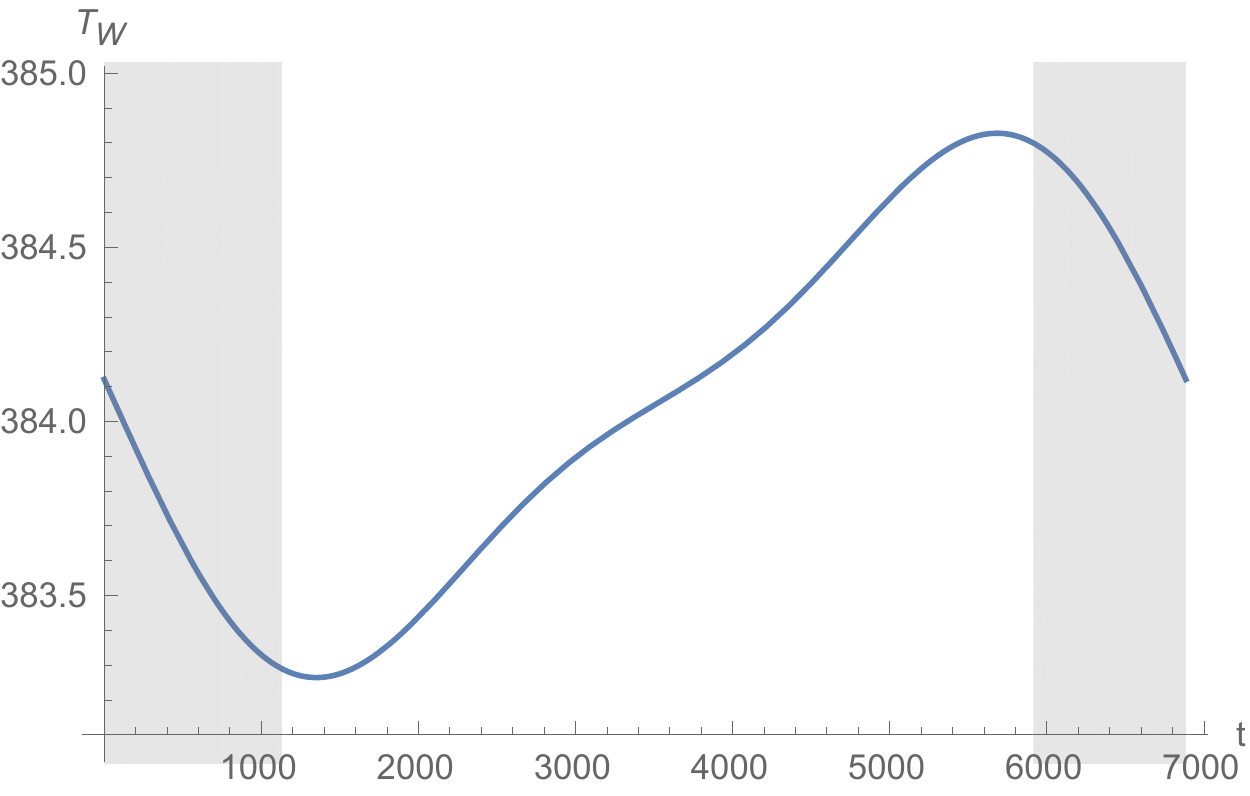} & \includegraphics[width=9cm]{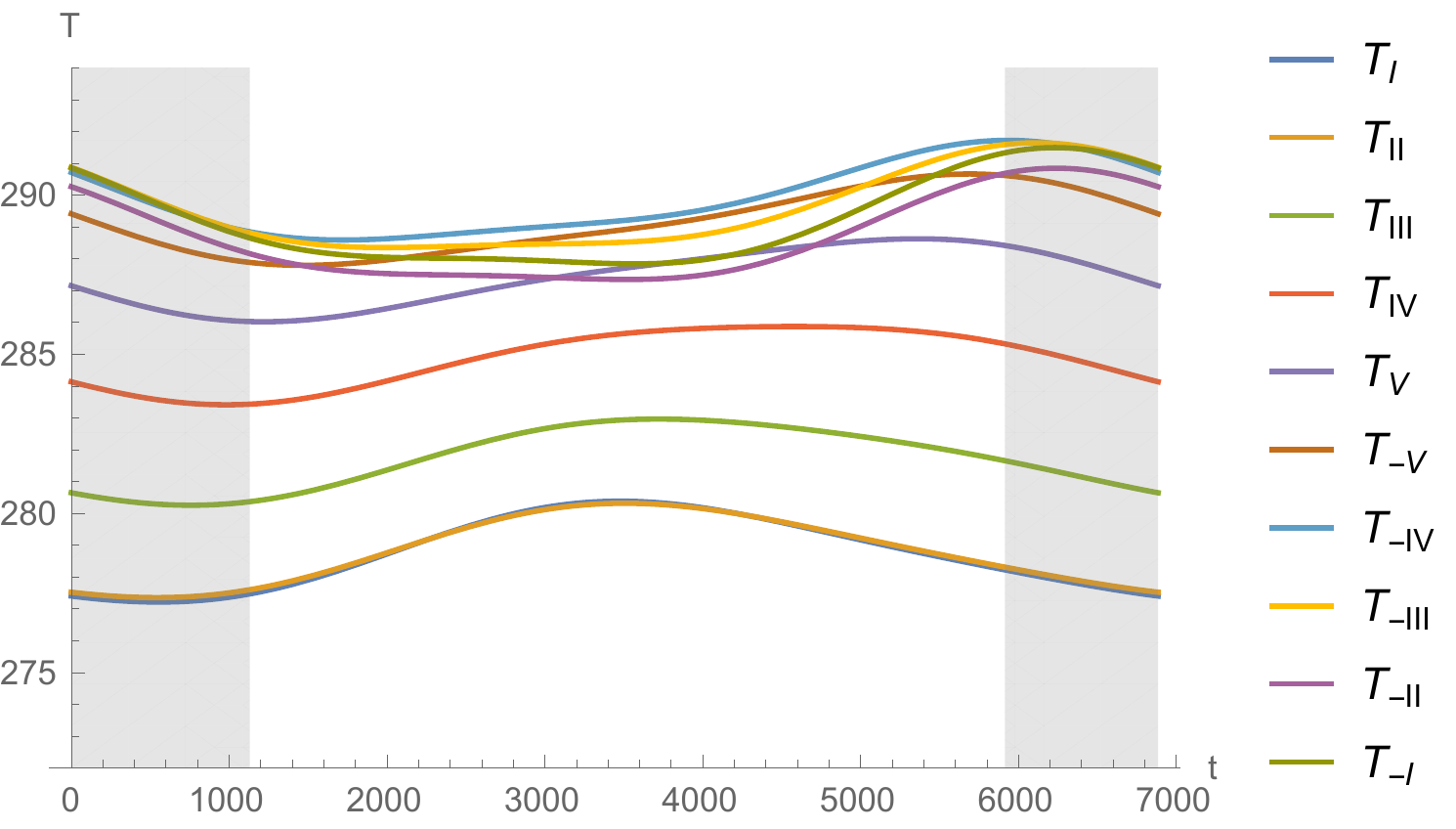}
\end{array}
$$
\caption{Top row: Plot of $T_{W}$ (left panel) and $T_{CCR}$ (right panel) (sum of the constant, first and second harmonics of the orbital frequency) over one period, on day 30, using the clean-glass absorptivity $\alpha_{gl,IR}=0.82$. The region colored gray is when the satellite is in the eclipse. The average along-track acceleration on this day is $-0.63 \ \mathrm{pm/s}^{2}$. For comparison, the average thermal drag over the 120-day period is $-0.59 \ \mathrm{pm/s}^{2}$. Bottom row: Plot of $T_{W}$ (left panel) and $T_{CCR}$ (right panel) (sum of the constant, first and second harmonics of the orbital frequency) over one period, on day 30, using the dirty-glass absorptivity $\alpha_{gl,IR}'=0.6$. The average along-track acceleration on this day is $-0.37 \ \mathrm{pm/s}^{2}$. For comparison, the average thermal drag over the 120-day period is $-0.36 \ \mathrm{pm/s}^{2}$. }
\label{TDay30}
\end{figure}
\begin{figure}
$$
\begin{array}{cc}
  \includegraphics[width=8cm]{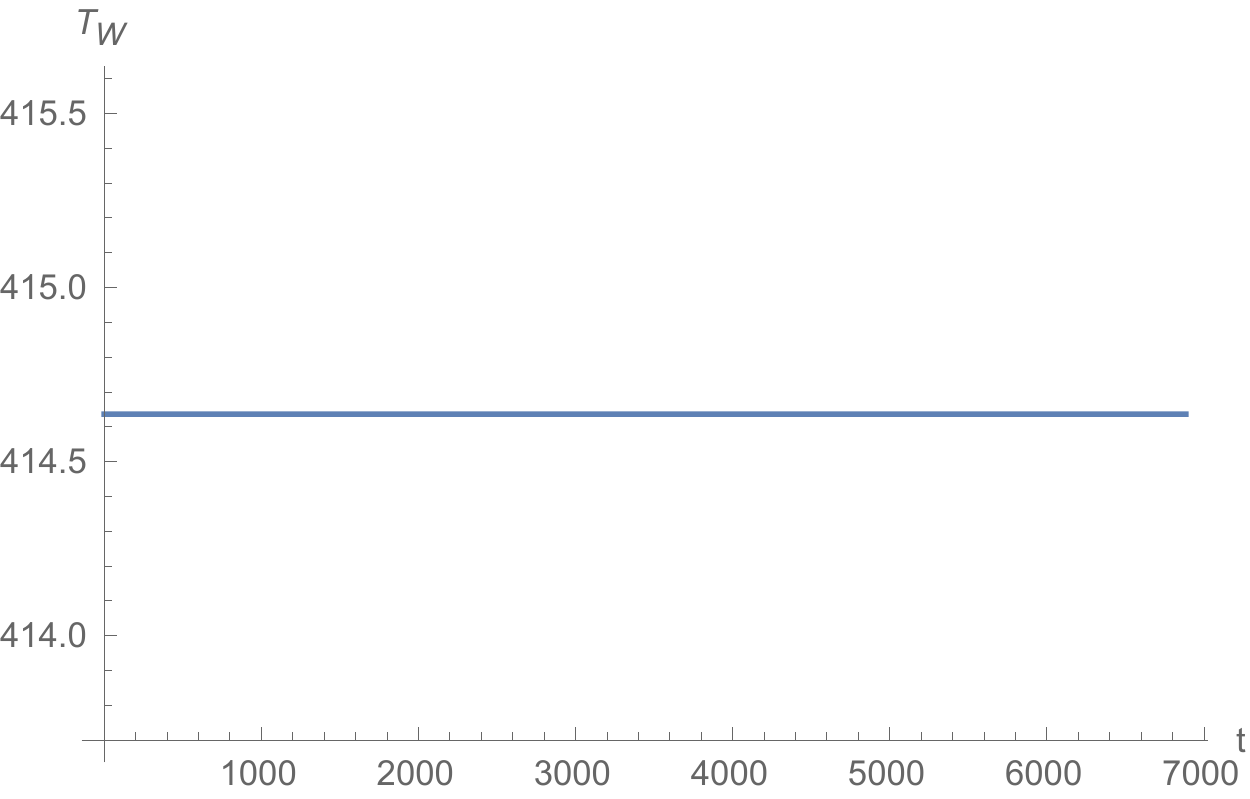} & \includegraphics[width=9cm]{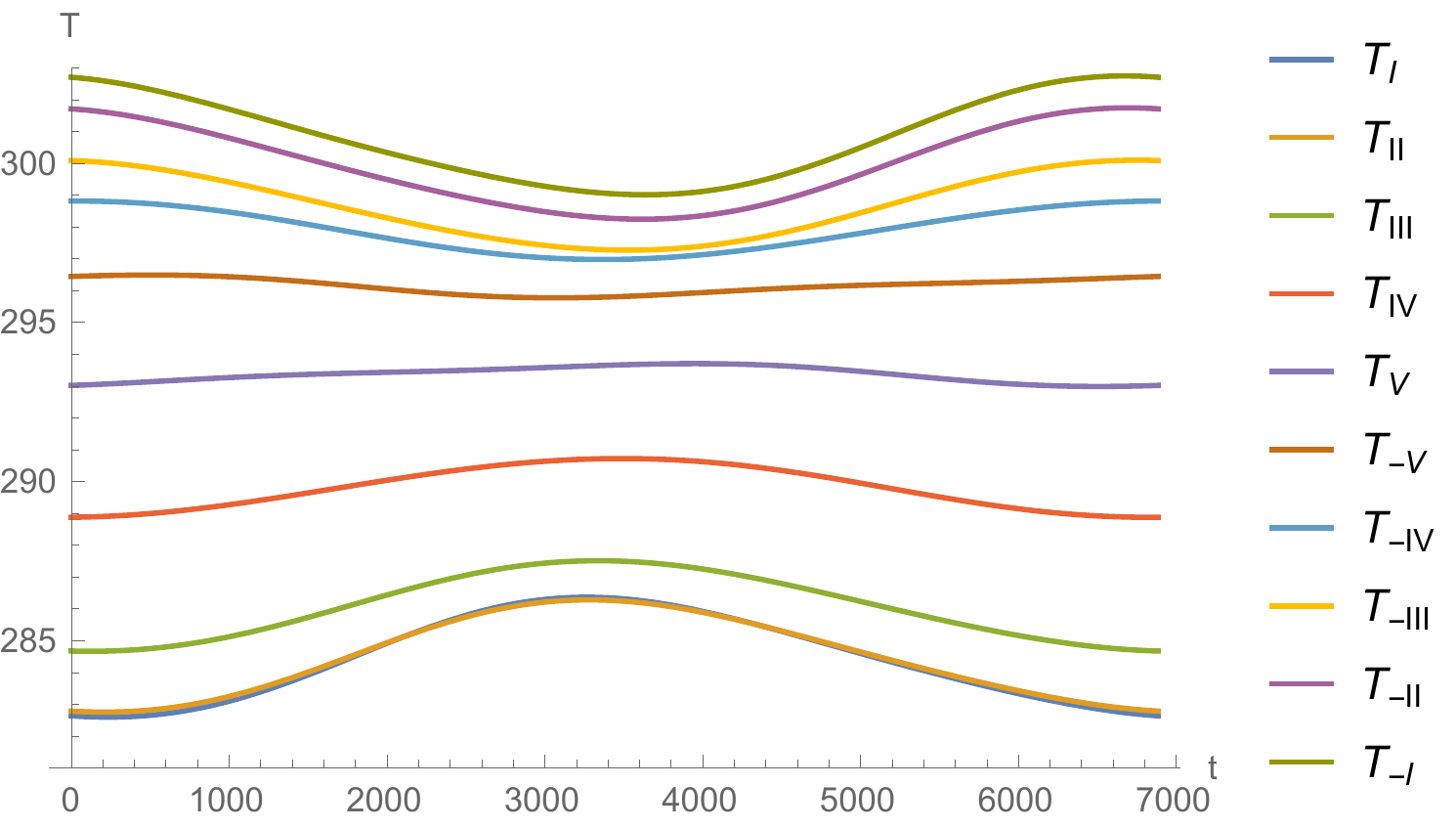} \\
    \includegraphics[width=8cm]{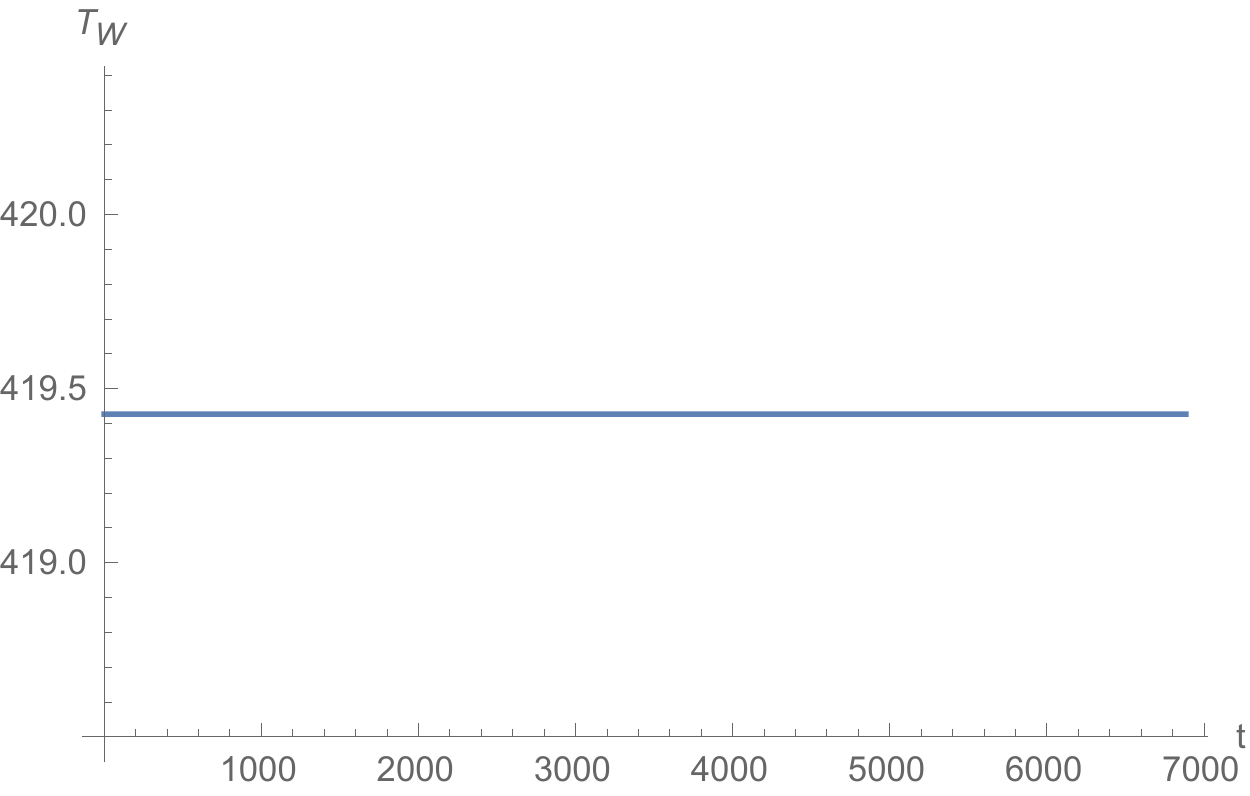} & \includegraphics[width=9cm]{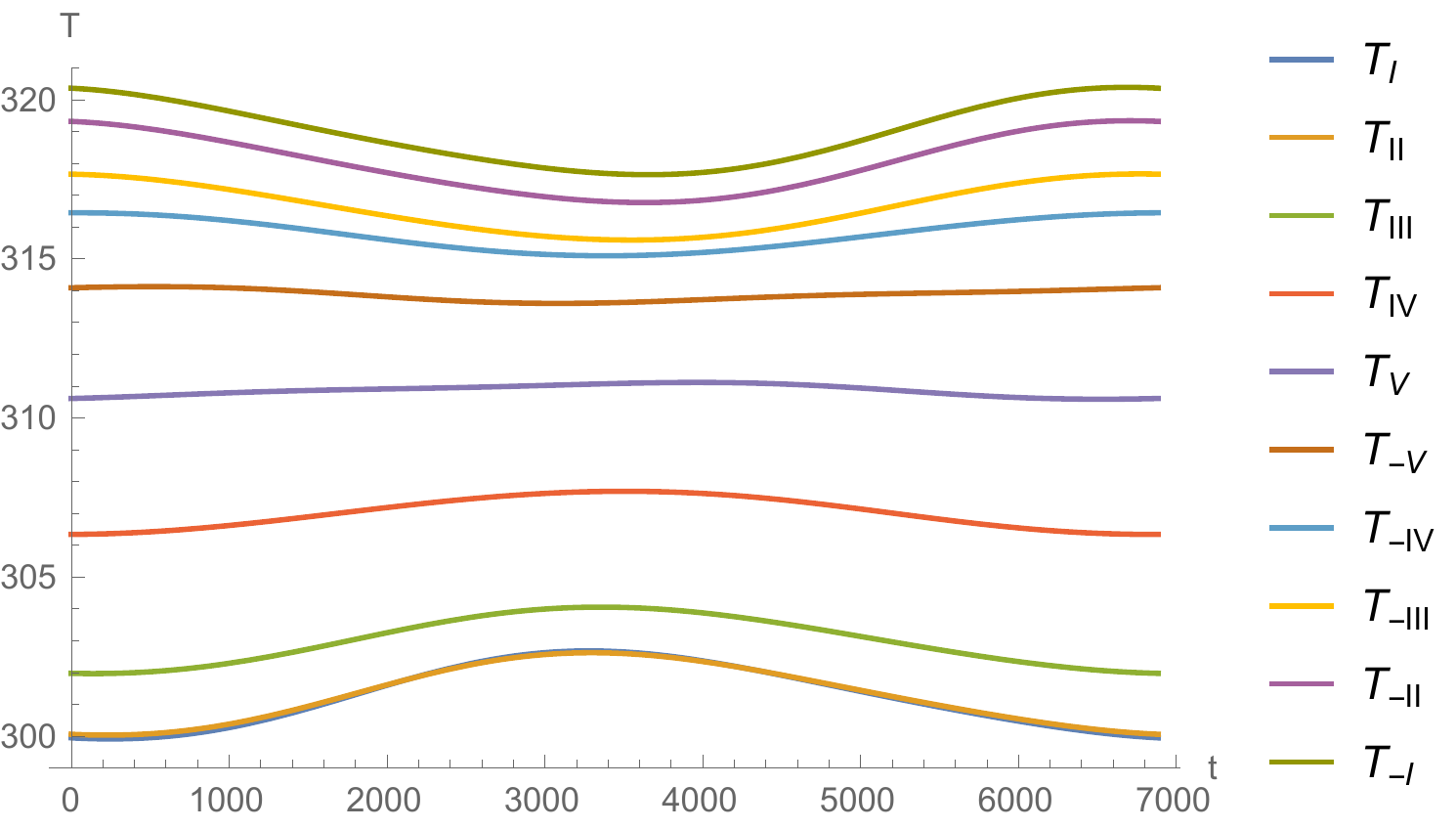}
\end{array}
$$
\caption{Top row: Plot of $T_{W}$ (left panel) and $T_{CCR}$ (right panel) (sum of the constant, first and second harmonics of the orbital frequency) over one period, on day 60, using the clean-glass absorptivity $\alpha_{gl,IR}=0.82$. There is no eclipse on this day, and as a result, the metal temperature is practically constant. The average along-track acceleration on this day is $-0.66 \ \mathrm{pm/s}^{2}$. For comparison, the average thermal drag over the 120-day period is $-0.59 \ \mathrm{pm/s}^{2}$. Bottom row: Plot of $T_{W}$ (left panel) and $T_{CCR}$ (right panel) (sum of the constant, first and second harmonics of the orbital frequency) over one period, on day 60, using the dirty-glass absorptivity $\alpha_{gl,IR}'=0.6$. The average along-track acceleration on this day is $-0.43 \ \mathrm{pm/s}^{2}$. For comparison, the average thermal drag over the 120-day period is $-0.36 \ \mathrm{pm/s}^{2}$. }
\label{TDay60}
\end{figure}
\begin{figure}
$$
\begin{array}{cc}
  \includegraphics[width=8cm]{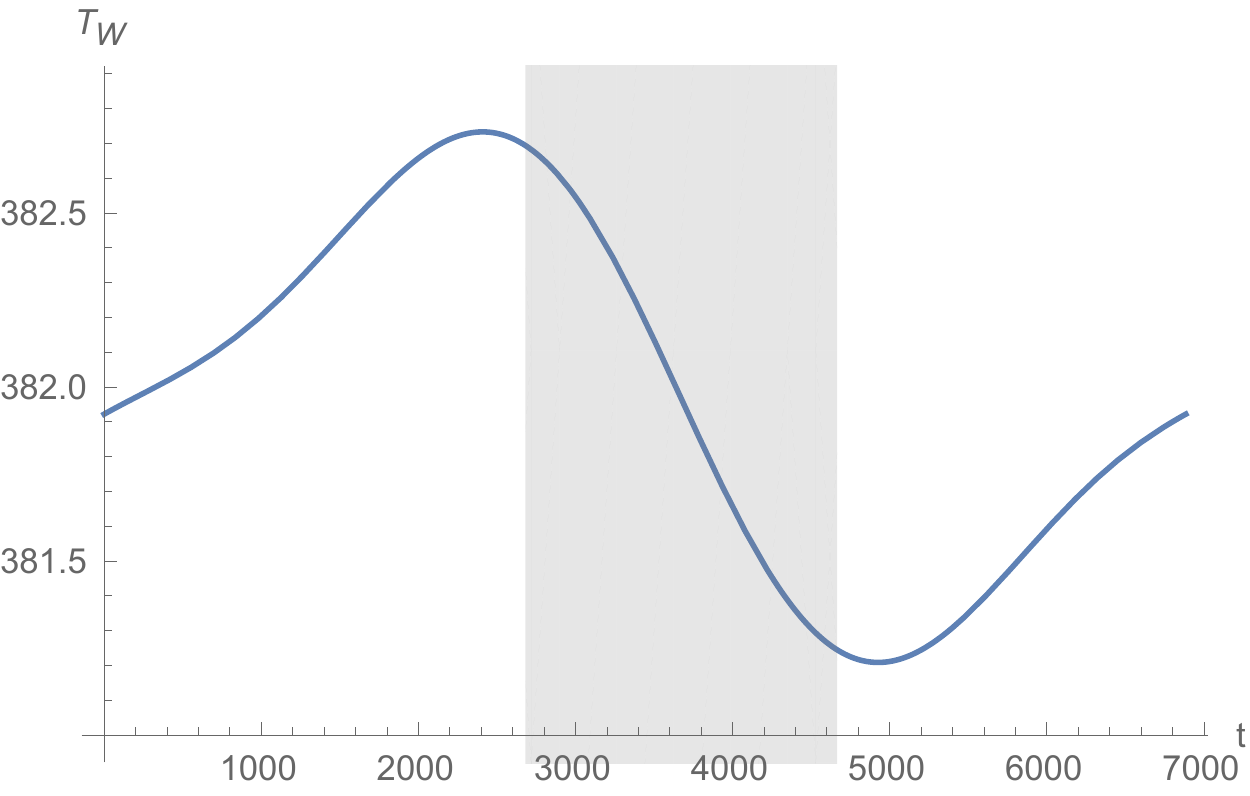} & \includegraphics[width=9cm]{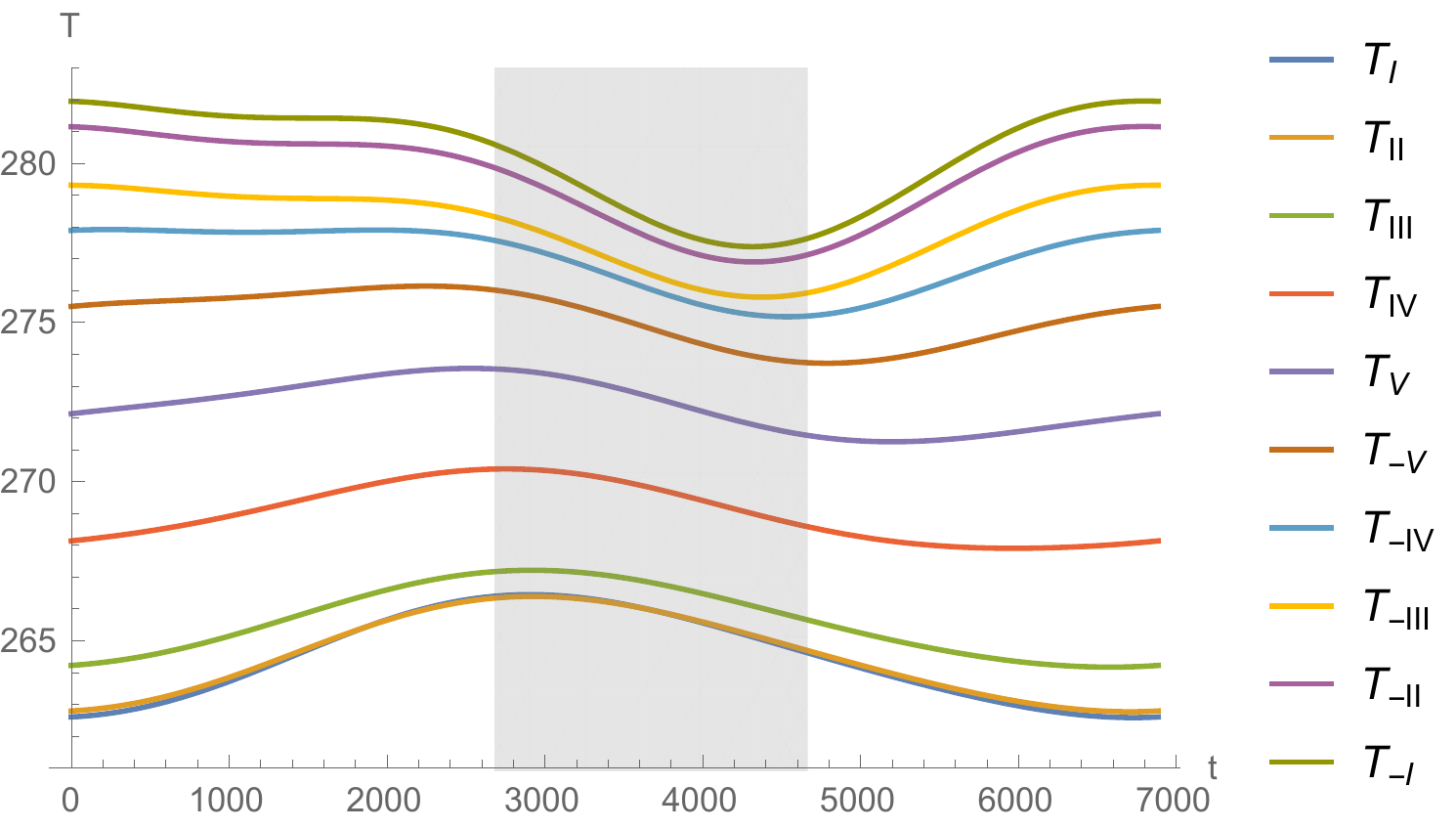} \\
  \includegraphics[width=8cm]{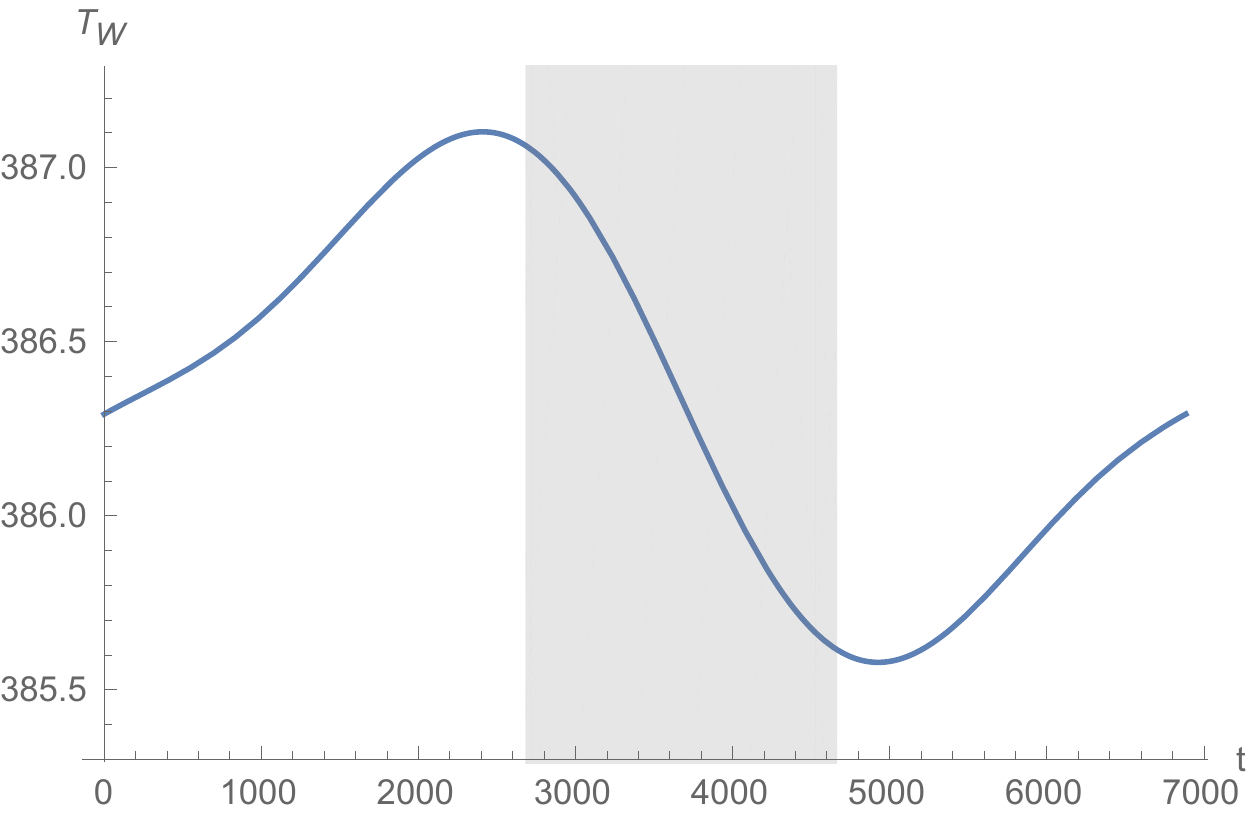} & \includegraphics[width=9cm]{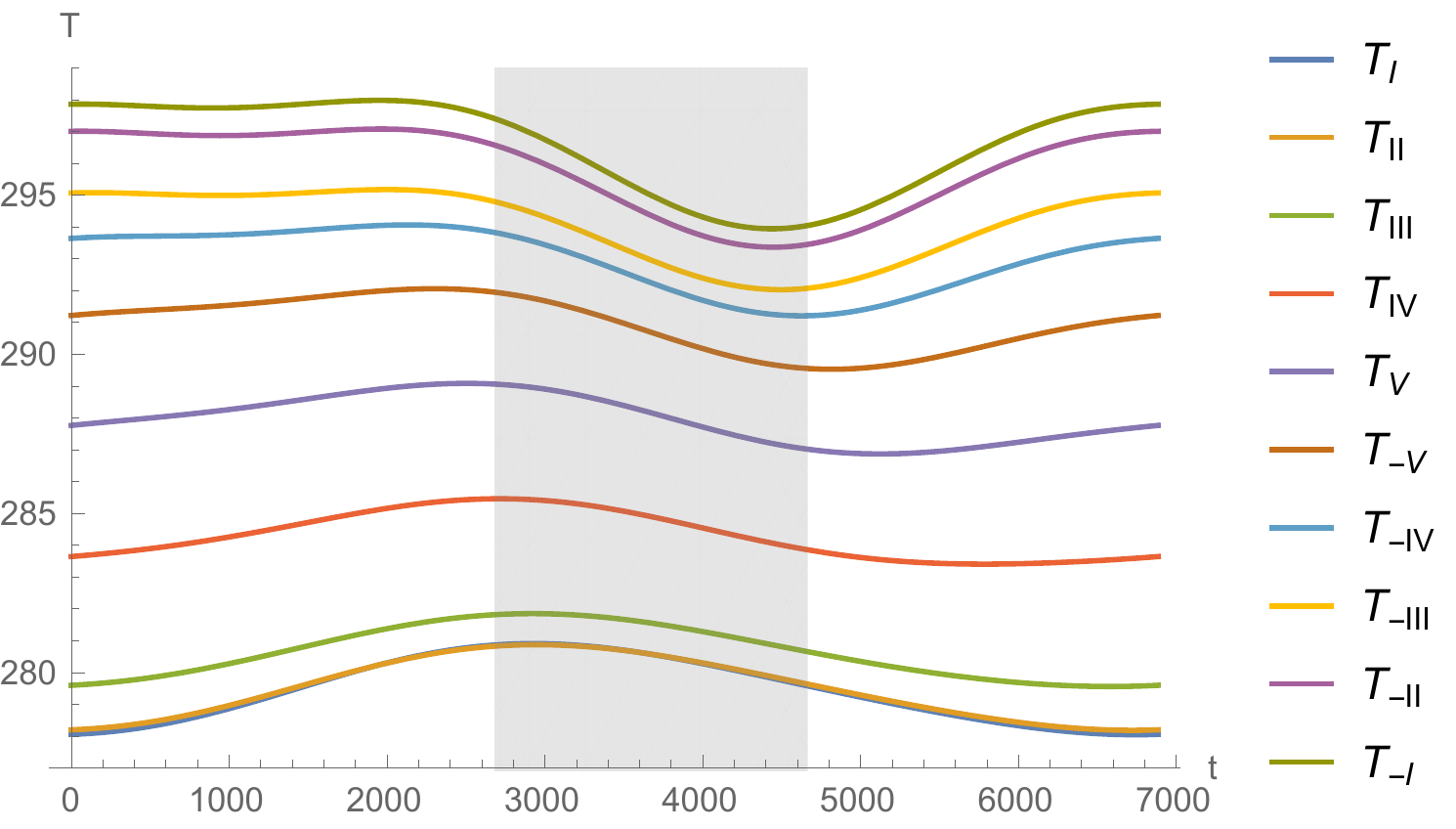}
\end{array}
$$
\caption{Top row: Plot of $T_{W}$ (left panel) and $T_{CCR}$ (right panel) (sum of the constant, first and second harmonics of the orbital frequency) over one period, on day 90, using the clean-glass absorptivity $\alpha_{gl,IR}=0.82$. The region colored gray is when the satellite is in the eclipse. The average along-track acceleration on this day is $-0.5 \ \mathrm{pm/s}^{2}$. For comparison, the average thermal drag over the 120-day period is $-0.59 \ \mathrm{pm/s}^{2}$. Bottom row: Plot of $T_{W}$ (left panel) and $T_{CCR}$ (right panel) (sum of the constant, first and second harmonics of the orbital frequency) over one period, on day 90, using the dirty-glass absorptivity $\alpha_{gl,IR}'=0.6$. The average along-track acceleration on this day is $-0.28 \ \mathrm{pm/s}^{2}$. For comparison, the average thermal drag over the first 120-day period is $-0.36 \ \mathrm{pm/s}^{2}$.}
\label{TDay90}
\end{figure}

\section{Conclusion}\label{Sec:Conclusion}
In this paper, we have demonstrated that the anomalous along-track acceleration of LARES can indeed be explained by a thermal effect: the anisotropic surface temperature of the satellite results in an anisotropic radiation pressure, and a net force imparted on the satellite. Our computation takes into account the detailed structure of the satellite. We linearize the Stefan-Boltzmann law, which allows for powerful Fourier series method. This is computationally much less intensive than a full numerical integration along the lines of \cite{Slabinski(1996)}. Our results are robust and consistent, given the small amplitude of temporal and spatial thermal fluctuations found.

\section{Acknowledgement}\label{Sec:Acknowledgement}
We thank Antonio Paolozzi, Ignazio Ciufolini, Victor Slabinski and Erricos Pavlis for useful comments. This material is supported by NASA under Grant No. NNX09AU86G. We also thank the Texas Cosmology Center, which is supported by the
College of Natural Sciences, the Department of Astronomy at the University of Texas at Austin, and the McDonald Observatory.

\appendix

\section{Radiative heat exchange}\label{App:ViewFactors}
Consider radiation exchange between $N$ surfaces labelled by $i$ ($i = 1, 2, \cdots, N$) inside an enclosure. The view factor $F_{ij}$ is defined to be the fraction of radiation leaving surface $i$ which is intercepted by surface $j$. The view factor can be computed by evaluating a double surface integral \cite{Incropera & Dewitt(1981)}:
\begin{equation}
F_{ij} = \frac{1}{A_{i}} \int_{A_{i}} \int_{A_{j}} \frac{\cos{\theta_{i}}\cos{\theta_{j}}}{\pi R^{2}} dA_{i} dA_{j}
\end{equation}
where $A_{i}$ is the surface element of surface $i$, $R$ is the length of the straight line connecting surface elements $dA_{i}$ with $dA_{j}$, and $\theta_{i}$ and $\theta_{j}$ are the angle formed by the normals of $dA_{i}$ and $dA_{j}$ with the line connecting the two elements. Fortunately, view factors can often be deduced without having to evaluate this integral. By interchanging $i$ and $j$ in the formula above, it follows that:
\begin{equation}\label{reciprocity}
A_{i}F_{ij} = A_{j}F_{ji}
\end{equation}
This is known as the reciprocity relation. Moreover, we have the summation rule
\begin{equation}\label{sumrule}
\sum_{j=1}^{N} F_{ij} = 1
\end{equation}
since radiation leaving surface $i$ must be intercepted by some surface lining the enclosure. Another observation which will prove useful to us is that if surface $i$ is convex, $F_{ii} = 0$ since radiation originating from $i$ cannot be intercepted by $i$. Once all view factors between pairs of surfaces are computed, we can compute the incident irradiance on surface $i$, denoted by $H_{i}$, due to radiation by all the surfaces in the enclosure:
\begin{equation}\label{incident}
H_{i} = \frac{1}{A_{i}} \sum_{j=1}^{N} F_{ji} J_{j} A_{j}
\end{equation}
where $J_{i}$ is the radiosity of surface $i$, that is, the sum of the emitted and reflected intensities:
\begin{equation}\label{radiosity}
J_{i} = \epsilon_{i} \sigma T_{i}^{4} + (1-\epsilon_{i})H_{i}
\end{equation}
Using the reciprocity relation (\ref{reciprocity}), (\ref{incident}) becomes:
\begin{equation}
H_{i} = \sum_{j=1}^{N} F_{ij} J_{j}
\end{equation}
Substituting into (\ref{radiosity}), we obtain a linear system of equations in the radiosities:
\begin{equation}\label{system}
J_{i} = \epsilon_{i} \sigma T_{i}^{4} + (1-\epsilon_{i}) \sum_{j=1}^{N} F_{ij} J_{j}
\end{equation}
Once the radiosities are solved for by the standard methods of linear algebra, we can form the net heat going into surface $i$, denoted by $P_{i}$, by taking the difference between the incoming and outgoing intensities:
\begin{equation}
P_{i} = A_{i}(H_{i}-J_{i})
\end{equation}
or, using (\ref{radiosity}) again,
\begin{equation}
P_{i} = \frac{A_{i}\epsilon_{i}}{1-\epsilon_{i}} (J_{i}-\sigma T_{i}^{4})
\end{equation}
Finally, we specialize to the case where the enclosure has only 2 isothermal elements, and we will assume moreover that element 1 is a convex surface. This scenario is useful for the purposes of a lumped-capacitance analysis of the CCR. The system \ref{system} in this case becomes:
\begin{equation}
J_{1} = \epsilon_{1} \sigma T_{1}^{4} + (1-\epsilon_{1}) J_{2}
\end{equation}
\begin{equation}
J_{2} = \epsilon_{2} \sigma T_{2}^{4} + (1-\epsilon_{2})[F_{21}J_{1}+(1-F_{21})J_{2}]
\end{equation}
where we used the fact that $F_{11}=0$, $F_{12}=1$ and equation (\ref{sumrule}). Solving for the net heat exchange, we find the net heat flux going into element 1 to be:
\begin{equation}\label{netheat}
P_{1} = A_{1}\epsilon_{eff}\sigma(T_{2}^{4}-T_{1}^{4})
\end{equation}
where we defined an effective emissivity $\epsilon_{eff}$:
\begin{equation}
\epsilon_{eff} = \bigg[\frac{1}{\epsilon_{1}} + \frac{1-\epsilon_{2}}{\epsilon_{2}}F_{21} \bigg]^{-1}
\end{equation}

\section{Numerical values of the constants}\label{App:Constants}
\begin{center}
\begin{tabular}{|l|c|r|}
  \hline
  Semi-major axis of LARES's orbit & $a$ & 7810 km \\
  Angular radius of Earth as viewed from LARES & $\alpha_{e}$ & 54.55 degrees \\
  Absorptivity/Emissivity of CCR in the IR (clean glass) \cite{Andres} & $\alpha_{gl,IR},\epsilon_{gl,IR}$ & 0.82 \\
  Absorptivity/Emissivity of CCR in the IR (dirty glass) & $\alpha_{gl,IR}',\epsilon_{gl,IR}'$ & 0.6 \\
  Absorptivity/Emissivity of CCR in the visible & $\alpha_{gl,vis},\epsilon_{gl,vis}$ & 0.15 \\
  Absorptivity/Emissivity of unoxidized tungsten in the IR at $500^{\circ}$C & $\alpha_{W,IR},\epsilon_{W,IR}$ & 0.07 \\
  Absorptivity/Emissivity of tungsten to sunlight & $\alpha_{W,vis}, \epsilon_{W,vis}$ & 0.45 \\
  Obliquity of the ecliptic & $\beta$ & 23.5 degrees \\
  Mean specific heat of Suprasil glass from $0^{\circ}$C to $500^{\circ}$C \cite{Heraeus} & $c_{gl}$ & 964 $\mathrm{J \cdot kg^{-1} \cdot K^{-1}}$ \\
  Specific heat capacity of W90 Ni6 Cu2-4 sintered tungsten \cite{AMS} & $c_{W}$ & 133.9 $\mathrm{J \cdot kg^{-1} \cdot K^{-1}}$ \\
  Distance from tip of CCR to bottom of cavity & $d$ & 5 $\mathrm{mm}$ \\
  Altitude of LARES & $h$ & $1540$ km \\
  Orbital inclination & $i$ & 70 degrees \\
  Thermal conductivity of Suprasil glass at $300^{\circ}$C \cite{Heraeus} & $\kappa_{gl}$ & 1.67 $\mathrm{W \cdot m^{-1} \cdot K^{-1}}$ \\
  Thermal conductivity of sintered tungsten THA-18N \cite{tungstenco} & $\kappa_{W}$ & 113 $\mathrm{W \cdot m^{-1} \cdot K^{-1}}$ \\
  Thermal diffusivity of tungsten & $\lambda_{W}$ & $4.688 \times 10^{-5}$ $\mathrm{m^{2} \cdot s^{-1}}$ \\
  Total mass of LARES & $m$ & 387.0 kg \\
  Mass of a CCR & $m_{CCR}$ & 0.03329 kg \\
  Earth IR flux per unit solid angle per unit receiver area & $N_{IR}$ & 71 $\mathrm{W \cdot m^{-2} \cdot sr^{-1}}$ \\
  Spinning frequency of LARES at launch & $\omega{(k=0)}$ & 0.546 $\mathrm{rad/s}$ \\
  Orbital frequency of LARES & $\omega_{o}$ & $9.13 \times 10^{-4} \mathrm{rad/s}$ \\
  Total solar irradiance & $\Phi$ & 1366 $\mathrm{W \cdot m^{-2}}$ \\
  Radius of CCR exterior surface & $R$ & 0.01905 $\mathrm{m}$ \\
  Radius of Earth's IR-emitting atmosphere & $R_{E}$ & 6407 km \\
  Radius of the satellite & $R_{sat}$ & 0.1820 $\mathrm{m}$ \\
  Mass density of sintered tungsten THA-18N \cite{tungstenco} & $\rho_{W}$ & 18000 $\mathrm{kg \cdot m^{-3}}$ \\
  Stefan-Boltzmann constant & $\sigma$ & $5.670 \times 10^{-8}$ $\mathrm{W \cdot m^{-2} \cdot K^{-4}}$ \\
  Period of LARES& $T$ & $114.7$ min \\
  \hline
\end{tabular}
\end{center}

\section{Analytical computation 1: the temperature of a stationary spinning sphere bathed in sunlight\label{App:SpinningSphere}}
In this appendix, we solve for the temperature inside a stationary, spinning metal sphere bathed in sunlight. We will orient the z-axis along the spin axis, and suppose that sunlight arrives at an angle $\beta_{0}$ from the spin axis. The heat equation reads:
\begin{equation}
\frac{\partial T}{\partial t} = \lambda_{W} \nabla^{2}T
\end{equation}
where $\lambda_{W}$ is the thermal diffusivity, defined as follows:
\begin{equation}
\lambda_{W} = \frac{\kappa_{W}}{c_{W}\rho_{W}}
\end{equation}
The boundary condition reads:
\begin{equation}
\alpha_{vis}I = \kappa \frac{\partial T}{\partial r}\bigg|_{r=R} + \epsilon_{W,IR}\sigma T(r=R)^{4}
\end{equation}
where $I$ is the intensity of radiation incident on the metal ball. It is straightforward to show that the intensity incident on the surface element at $P = (\theta,\phi)$ takes the form:
\begin{equation}\label{incidentintensity}
I(t) = \Phi \cos{\gamma{(t)}} \Theta{(\cos{\gamma{(t)}})}
\end{equation}
where $\Phi$ is the solar constant, $\Theta$ is the Heaviside step function, $\gamma$ is the (time-dependent) angle formed by the position vector of $P$ with the direction of the Sun, and
\begin{equation}
\cos{(\gamma{(t)})} = \cos{\beta_{0}}\cos{\theta} + \sin{\beta_{0}}\sin{\theta}\cos{(\phi-\omega t)}
\end{equation}
Notice that the boundary condition is nonlinear in $T$ due to the Stefan-Boltzmann term. As in the rest of the paper, we linearlize the boundary condition by letting $T = T_{0} + \Delta T$ where $T_{0}$ is the average temperature of the sphere, and $\Delta T$ is some small deviation from the average. The boundary value problem becomes:
\begin{equation}
\frac{\partial \Delta T}{\partial t} = \lambda_{W} \nabla^{2} \Delta T
\end{equation}
\begin{equation}
\alpha_{vis} I -\epsilon \sigma T_{0}^{4} = \kappa \frac{\partial \Delta T}{\partial r} \bigg|_{r=R} + 4 \epsilon \sigma T_{0}^{3} \Delta T (r=R)
\end{equation}
Thanks to time periodicity, we can expand the radiant intensity in a Fourier series:
\begin{equation}
I(\theta,\phi, t) = \sum_{n=-\infty}^{\infty} I_{n}(\theta,\phi) e^{in\omega t}
\end{equation}
where $\omega$ is the spin frequency (for simplicity, we are not considering any orbital motion in this appendix). The Fourier modes are determined by:
\begin{equation}
I_{n}(\theta,\phi) = \frac{\omega}{2\pi} \int_{-\pi/\omega}^{\pi/\omega} I(\theta,\phi,t) e^{-in\omega t} dt
\end{equation}
To evaluate, we distinguish between 3 regions on the sphere. First, consider the case when $0 \leq \theta \leq \frac{\pi}{2}-\beta_{0}$. In this case, the surface element is permanently illuminated, and
\begin{equation}
I_{n}(\theta,\phi) = \frac{\omega\Phi}{2\pi} \int_{-\pi/\omega}^{\pi/\omega} (\cos{\beta_{0}}\cos{\theta}+\sin{\beta_{0}}\sin{\theta}\cos{(\phi-\omega t)})e^{-in\omega t} dt
\end{equation}
In this case, only the terms with $n=0$ and $n = \pm 1$ contribute:
\begin{equation}
I_{0}(\theta,\phi) = \Phi\cos{\beta_{0}}\cos{\theta}
\end{equation}
\begin{equation}
I_{\pm1}(\theta,\phi) = \frac{\Phi}{2}\sin{\beta_{0}}\sin{\theta}e^{\mp i \phi} 
\end{equation}
Next, consider the case $\frac{\pi}{2} - \beta_{0} \leq \theta \leq \frac{\pi}{2} + \beta_{0}$. In this case, the surface element is only illuminated part of the time:
\begin{equation}
I_{n}(\theta,\phi) = \frac{\omega\Phi}{2\pi} \int_{(\phi-\phi_{0})/\omega}^{(\phi+\phi_{0})/\omega} (\cos{\beta_{0}}\cos{\theta}+\sin{\beta_{0}}\sin{\theta}\cos{(\phi-\omega t)})e^{-in\omega t} dt
\end{equation}
where
\begin{equation}
\phi_{0}(\theta,\beta_{0}) = \arccos{(-\cot{\beta_{0}}\cot{\theta})}
\end{equation}
In this case, infinitely many values of $n$ contribute, and the first few are:
\begin{equation}
I_{0}(\theta,\phi) = \frac{\Phi}{\pi}(\phi_{0} \cos{\beta_{0}}\cos{\theta} + \sin{\phi_{0}}\sin{\beta_{0}}\sin{\theta})
\end{equation}
\begin{equation}
I_{\pm 1}(\theta,\phi) = \frac{\Phi}{\pi}e^{\mp i \phi}\bigg[\sin{\phi_{0}}\cos{\beta_{0}}\cos{\theta}+\frac{1}{2}(\phi_{0}+\cos{\phi_{0}}\sin{\phi_{0}})\sin{\beta_{0}}\sin{\theta}\bigg]
\end{equation}
Finally, for the region $\frac{\pi}{2}+\beta_{0} \leq \theta \leq \pi$, the surface element remains in the dark at all time, and $I_{n}(\theta,\phi)=0$ for all $n$. Notice that $I_{0}$ is nothing but the average intensity over a revolution. As expected, $I_{0}$ is not a function of $\phi$ but only of $\theta$.

We also expand the temperature in a Fourier series (with the same frequency $\omega$):
\begin{equation}
\Delta T (r, \theta, \phi, t) = \sum_{n=-\infty}^{\infty} \Delta T_{n}(r,\theta,\phi) e^{in\omega t}
\end{equation}
Since the left-hand side must be real, we have $(\Delta T_{n})^{*} = \Delta T_{-n}$, and
\begin{equation}
\Delta T(r,\theta,\phi,t) = \sum_{n=0}^{\infty} 2 Re\{\Delta T_{n} e^{in\omega t}\}
\end{equation}
Substituting into the heat equation, we find that each mode $\Delta T_{n}$ satisfies the Helmholtz equation:
\begin{equation}
(\nabla^{2}+k_{n}^{2})\Delta T_{n} = 0
\end{equation}
where
\begin{equation}
k_{n}^{2} = -\frac{i n \omega}{\lambda_{W}}
\end{equation}
Unlike the usual Helmholtz equation, however, the square of the ``wavenumber'' $k_{n}$ in this case is imaginary. For the special case $n=0$, we have $k_{0}=0$ and the Laplace equation:
\begin{equation}
\nabla^{2} \Delta T_{0} = 0
\end{equation}
The general solution is:
\begin{equation}
\Delta T_{0}(r,\theta,\phi) = \sum_{l=0}^{\infty} \sum_{m=-l}^{l}(A_{lm}r^{l}+B_{lm}r^{-l-1}) Y_{l}^{m}(\theta,\phi)
\end{equation}
Since the temperature has to be bounded at $r=0$, we set $B_{l}=0$. Also, we will start the sum over $l$ at $l=1$ since the $l=m=0$ term is constant and can be absorbed into the average temperature $T_{0}$. For all other values of n, $k_{n}$ is nonzero and the general solution to Helmholtz equation is:
\begin{equation}
\Delta T_{n}(r,\theta,\phi) = \sum_{l=0}^{\infty} \sum_{m=-l}^{l} (a_{lmn}j_{l}(k_{n}r)+b_{lmn}y_{l}(k_{n}r)) Y_{l}^{m}(\theta,\phi)
\end{equation}
where $j_{l}$ and $y_{l}$ are the spherical Bessel functions and $Y_{l}^{m}$ is the spherical harmonics. Requiring boundedness at $r=0$ again, we set $b_{lmn} = 0$. The general solution for $\Delta T$ is then:
\begin{equation}
\Delta T(r,\theta,\phi,t) = \sum_{l,m} A_{lm}r^{l}Y_{l}^{m}(\theta,\phi) + \sum_{n \in \mathrm{N}} \sum_{l,m} 2\mathrm{Re}{\bigg[a_{lmn} j_{l}{\bigg(\sqrt{n\omega/\lambda_{W}}r e^{3\pi i/4}\bigg)}Y_{l}^{m}(\theta,\phi)e^{in\omega t}\bigg]}
\end{equation}
We will next compute the first few coefficients of this series. By substituting the general solution into the boundary condition, we obtain for $n=0$:
\begin{equation}
\alpha_{vis} I_{0}(\theta,\phi) - \epsilon \sigma T_{0}^{4} = \sum_{l,m} A_{lm} (\kappa l R^{l-1} + 4\epsilon \sigma T_{0}^{3}R^{l}) Y_{l}^{m}(\theta,\phi)
\end{equation}
and for $n \neq 0$:
\begin{equation}
\alpha_{vis} I_n(\theta,\phi) = \sum_{l=0}^{\infty} \sum_{m=-l}^{l} a_{lmn} (\kappa k_{n} j_{l}'(k_{n}R)+4\epsilon\sigma T_{0}^{3}j_{l}(k_{n}R)) Y_{l}^{m}(\theta,\phi)
\end{equation}
where the derivative of the spherical Bessel functions can be expressed in terms of the spherical Bessel functions as:
\begin{equation}
(2l+1)j_{l}'{(z)} = l j_{l-1}{(z)} - (l+1)j_{l+1}{(z)}
\end{equation}
It remains to expand $I_{n}(\theta,\phi)$ in the spherical coordinates and equate coefficients:
\begin{equation}
I_{n}(\theta,\phi) = \sum_{l,m} I_{nlm} Y_{l}^{m}(\theta,\phi)
\end{equation}
where, using orthonormality of the spherical harmonics, we find:
\begin{equation}
I_{nlm} = \int_{\phi =0}^{2\pi} \int_{\theta=0}^{\pi} I_{n}(\theta,\phi) Y_{l}^{m*}(\theta,\phi) \sin{\theta} d\theta d\phi
\end{equation}
A remarkable simplification occurs if we consider the integral over $\phi$:
\begin{equation}
I_{nlm} \propto \int_{0}^{2\pi} e^{\mp i(n+m)\phi} d\phi
\end{equation}
it is easily seen that $I_{nlm}$ vanishes unless $m=-n$. Thus, it is enough to compute the coefficients $I_{nl-n}$ for $l \geq n$. In particular, among the coefficients $A_{lm}$, only those of the form $A_{l0}$ contribute. As a result, this series has no $\phi$-dependence and can be rewritten as a series in the Legendre polynomials. Among the coefficients $a_{nlm}$, only those of the form $a_{nl-n}$ where $l \geq n$ are nonzero.

We now proceed to compute the first few $I_{nlm}$. For $n=0$, the first coefficient is $I_{000}$ and it is given by:
\begin{eqnarray}
I_{000} &=& \frac{\Phi \sqrt{\pi}\cos^{3}{\beta}}{2} + \frac{\Phi}{\sqrt{\pi}}\cos{\beta} \int_{\frac{\pi}{2}-\beta}^{\frac{\pi}{2}+\beta}\arccos{(-\cot{\beta}\cot{\theta})}\cos{\theta}\sin{\theta}d\theta \nonumber \\
&+& \frac{\Phi}{\sqrt{\pi}}\sin{\beta} \int_{\frac{\pi}{2}-\beta}^{\frac{\pi}{2}+\beta} \sqrt{1-\cot^{2}{\beta}\cot^{2}{\theta}}\sin^{2}{\theta}d\theta
\end{eqnarray}
With the change of variable $u = -\cot{\beta}\cot{\theta}$, these integrals can be evaluated and the final result is:
\begin{equation}
I_{000} = \frac{\Phi\sqrt{\pi}}{2}
\end{equation}
In particular, $I_{000}$ is independent of both $\omega$ and $\beta$. This is expected since it is a time-averaged, spatially averaged intensity flux. The average temperature $T_{0}$ will be found from the fact that $A_{00}=0$:
\begin{equation}
T_{0} = \bigg(\frac{\alpha_{vis}I_{000}}{\epsilon_{IR}\sigma 2\sqrt{\pi}}\bigg)^{1/4} = 443.6\ \mathrm{K}
\end{equation}
For the next coefficients $I_{nlm}$, we will only evaluate the integrals in for the particular case $\beta_{0}=0$:
\begin{equation}
I_{010} = \Phi \sqrt{\frac{\pi}{3}}
\end{equation}
\begin{equation}
I_{020} = \frac{\sqrt{5\pi}}{8}\Phi
\end{equation}
The time-independent piece of the temperature variation in then found to be:
\begin{equation}
\Delta T_{0}{(r,\theta)} = \frac{\alpha_{vis}\Phi}{2(\kappa+4\epsilon_{W,IR}\sigma T_{0}^{3}R_{sat})} r\cos{\theta} + \frac{5\alpha_{vis}\Phi}{32(2R\kappa+4\epsilon_{W,IR}\sigma T_{0}^{3}R^{2})} r^{2}(3\cos^{2}{\theta}-1)
\end{equation}
Or, numerically:
\begin{equation}\label{DeltaTlinearized}
\Delta T_{0}(r,\theta) = 0.494 \frac{r}{R_{sat}}\cos{\theta} + 0.077 \left(\frac{r}{R_{sat}}\right)^{2}(3\cos^{2}{\theta}-1)
\end{equation}
Next, we compute the coefficients $a_{lmn}$ of the time-dependent piece. For $n=1$ (and general $\beta_{0}$), the lowest order term is:
\begin{eqnarray}
I_{11-1} &=& \frac{\Phi}{24}\sqrt{\frac{3\pi}{2}}\sin{\beta_{0}}(8-9\sin{\beta_{0}}-\sin{3\beta_{0}}) \\
&+& \frac{\Phi}{2}\sqrt{\frac{3}{2\pi}}\sin{\beta_{0}} \int_{\frac{\pi}{2}-\beta_{0}}^{\frac{\pi}{2}+\beta_{0}}\arccos{(-\cot{\beta_{0}}\cot{\theta})}\sin^{3}{\theta}d\theta \nonumber
\end{eqnarray}
As a consistency check, notice that if we set $\beta_{0}=0$ in the expression above, we find $I_{11-1}=0$. This makes sense since, for $\beta_{0}=0$, sunlight is incident from the same direction as the spin axis, and the angle of incidence for each CCR row is time-independent. Thus one can expect the coefficient above to vanish.

\section{Analytical computation 2: Going beyond the linearized boundary condition}\label{App:ClebschGordan}
Throughout the paper, we have systematically linearized the radiative boundary condition. In this appendix, we develop an approximation scheme to go beyond the linearized boundary condition. To avoid complications due to time-dependence, we consider a non-spinning sphere bathing in sunlight. Without loss of generality, let sunlight arrive from the direction $\theta=0$. Inside the ball, the Laplace equation for the metal is satisfied:
\begin{equation}
\nabla^{2} T = 0
\end{equation}
At the surface, the following boundary condition is satisfied:
\begin{equation}\label{bc}
\kappa_{W}\frac{\partial T}{\partial r}\bigg|_{(R,\theta)} + \epsilon_{W,IR} \sigma T{(R,\theta)}^{4} = \alpha_{W,vis}\Phi\cos{\theta}\Theta{\left(\frac{\pi}{2}-\theta\right)}
\end{equation}
Since the temperature is independent of $\phi$ by symmetry, we can expand in the Legendre polynomials:
\begin{equation}
T{(r,\theta)} = \sum_{l=0}^{\infty} \frac{A_{l}}{\sqrt{2l+1}}r^{l}Y_{l}^{0}{(\theta)}
\end{equation}
Explicitly, to second order this reads:
\begin{equation}
T{(r,\theta)} = A_{0}Y_{0}^{0}{(\theta)} + \frac{A_{1}}{\sqrt{3}}rY_{1}^{0}{(\theta)} + \frac{A_{2}}{\sqrt{5}}r^{2}Y_{2}^{0}{(\theta)}
\end{equation}
The idea now is to expand both sides of the boundary condition (\ref{bc}) as a series in the Legendre polynomials, without linearizing the $T^{4}$ term. To do this, we will have to expand a product of spherical harmonics in terms of the spherical harmonics themselves. This is provided by the following formula:
\begin{equation}
Y_{l_{1}}^{m_{1}}{(\theta,\phi)}Y_{l_{2}}^{m_{2}}{(\theta,\phi)} = \sum_{l=|l_{1}-l_{2}|}^{l_{1}+l_{2}} \sqrt{\frac{(2l_{1}+1)(2l_{2}+1)}{4\pi(2l+1)}} \left<l0|l_{1}l_{2}00\right>\left<l(m_{1}+m_{2})|l_{1}l_{2}m_{1}m_{2}\right> Y_{l}^{m_{1}+m_{2}}{(\theta,\phi)}
\end{equation}
where $\left<lm|l_{1}l_{2}m_{1}m_{2}\right>$ are known as the Clebsch-Gordan coefficients. In particular,
\begin{equation}
Y_{1_{1}}^{0}{(\theta)}Y_{l_{2}}^{0}{(\theta)} = \sum_{l=|l_{1}-l_{2}|}^{l_{1}+l_{2}} \sqrt{\frac{(2l_{1}+1)(2l_{2}+1)}{4\pi(2l+1)}} \left<l0|l_{1}l_{2}00\right>^{2} Y_{l}^{0}{(\theta)}
\end{equation}
The expansion for $T^{4}$ is then:
\begin{eqnarray}
T^{4}{(r,\theta)} = \sum_{l_{1},l_{2},l_{1}',l_{2}'=0}^{\infty} \sum_{l=|l_{1}-l_{2}|}^{l_{1}+l_{2}} \sum_{l'=|l_{1}'-l_{2}'|}^{l_{1}'+l_{2}'}\sum_{l''=|l-l'|}^{l+l'} \frac{A_{l_{1}}A_{l_{2}}A_{l_{1}'}A_{l_{2}'}}{(4\pi)^{3/2}\sqrt{2l''+1}}r^{l_{1}+l_{2}+l_{1}'+l_{2}'} \nonumber \\
\times \left<l0|l_{1}l_{2}00\right>^{2} \left<l'0|l_{1}'l_{2}'00\right>^{2} \left<l''0|ll'00\right>^{2} Y_{l''}^{0}{(\theta)}
\end{eqnarray}
We will refer to the coefficient of $r^{n}Y_{l}^{m}$ in the expansion above as $C_{lmn}$. Suppose we work to second order, and compute all $C_{lmn}$ with $n \leq 2$ and $l \leq 2$. For example, let us compute $C_{000}$ first. To do this, we have to consider all possible values of $l_{1}$, $l_{2}$, $l_{1}'$, $l_{2}'$, $l$, $l'$ and $l''$ such that $l_{1}+l_{2}+l_{1}'+l_{2}'=0$ and $l''=0$. But there is only one possible assignment of values: $l_{1}=l_{2}=l_{1}'=l_{2}'=l=l'=l''=0$. Therefore:
\begin{equation}
C_{000} = \frac{A_{0}^{4}}{(4\pi)^{3/2}} \left<00|0000\right>^{6} = \frac{A_{0}^{4}}{(4\pi)^{3/2}}
\end{equation}
Proceeding similarly for the other coefficients, we find the following expansion for $T^{4}$ (to second order in $n$ and $l$):
\begin{equation}
T^{4}{(r,\theta)} = \frac{1}{(4\pi)^{3/2}}\left(A_{0}^{4} + \frac{4}{3}A_{0}^{2}A_{2}^{2}r^{2}\right)Y_{0}^{0}{(\theta)} + \frac{A_{1}A_{0}^{3}}{\sqrt{3}(4\pi)^{3/2}}rY_{1}^{0}{(\theta)} + \frac{8}{3\sqrt{5}}\frac{A_{0}^{2}A_{1}^{2}}{(4\pi)^{3/2}}r^{2}Y_{2}^{0}{(\theta)}
\end{equation}
The right-hand side of (\ref{bc}) can also be expanded as follows:
\begin{equation}
\alpha_{W,IR}\Phi\cos{\theta}\Theta{\left(\frac{\pi}{2}-\theta\right)} = \alpha_{W,IR}\Phi\left(\frac{\sqrt{\pi}}{2}Y_{0}^{0} + \sqrt{\frac{\pi}{3}}Y_{1}^{0} + \frac{\sqrt{5\pi}}{8}Y_{2}^{0} \right)
\end{equation}
From the boundary condition then, we obtain a system of 3 nonlinear equations in 3 unknowns ($A_{0}$, $A_{1}$ and $A_{2}$). Solving for these coefficients, we find:
\begin{equation}
T{(r,\theta)} = 443.6 +0.495\frac{r}{R_{sat}}\cos{\theta} + 0.077\left(\frac{r}{R_{sat}}\right)^{2}(3\cos^{2}{\theta}-1)
\end{equation}
Comparing with equation (\ref{DeltaTlinearized}) of the previous section, we find an almost identical result. Thus, the nonlinearity in the boundary condition really does not matter much.

\end{document}